\documentclass[a4paper,11pt]{article}
\pdfoutput=1

\usepackage{jcappub}
                     
\usepackage{aas_macros} 

\usepackage[T1]{fontenc}
\usepackage[utf8]{inputenc}

\usepackage{titlesec}

\usepackage[english]{babel}
\usepackage{amsmath}
\usepackage{amssymb}
\usepackage{MnSymbol}
\usepackage{nicefrac}
\usepackage{amsfonts}
\usepackage{dsfont}
\usepackage[justification=centering]{subcaption,ragged2e}
\usepackage{mathtools}

\usepackage{natbib}
\usepackage{hyperref}

\usepackage[markup=nocolor]{changes}

\usepackage{graphicx}

\usepackage[page,header]{appendix}
\usepackage{titletoc}

\usepackage{dcolumn}
\usepackage{bm}

\usepackage{color}
\usepackage{xcolor}
\newcommand{\bea}{\begin{eqnarray}}
\newcommand{\eea}{\end{eqnarray}}
\newcommand{\be}{\begin{equation}}
\newcommand{\ee}{\end{equation}}

\DeclareRobustCommand{\looongrightarrow}{
  \DOTSB\relbar\joinrel\relbar\joinrel\relbar\joinrel\rightarrow
}

\titleclass{\subsubsubsection}{straight}[\subsection]

\newcounter{subsubsubsection}[subsubsection]
\renewcommand\thesubsubsubsection{\thesubsubsection.\arabic{subsubsubsection}}

\titleformat{\subsubsubsection}
  {\normalfont\normalsize\bfseries}{\thesubsubsubsection}{1em}{}
\titlespacing*{\subsubsubsection}
{0pt}{3.25ex plus 1ex minus .2ex}{1.5ex plus .2ex}

\makeatletter
\renewcommand\paragraph{\@startsection{paragraph}{5}{\z@}%
  {3.25ex \@plus1ex \@minus.2ex}%
  {-1em}%
  {\normalfont\normalsize\bfseries}}
\renewcommand\subparagraph{\@startsection{subparagraph}{6}{\parindent}%
  {3.25ex \@plus1ex \@minus .2ex}%
  {-1em}%
  {\normalfont\normalsize\bfseries}}
\def\toclevel@subsubsubsection{4}
\def\toclevel@paragraph{5}
\def\toclevel@paragraph{6}
\def\l@subsubsubsection{\@dottedtocline{4}{7em}{4em}}
\def\l@paragraph{\@dottedtocline{5}{10em}{5em}}
\def\l@subparagraph{\@dottedtocline{6}{14em}{6em}}
\makeatother

\title{The principled-parameterized approach to gravitational collapse}

\author[a]{H\'elo\"ise Delaporte,}
\author[a]{Astrid Eichhorn}

\affiliation[a]{CP3-Origins, University of Southern Denmark,
\\
Campusvej 55, DK-5230 Odense M, Denmark}

\emailAdd{eichhorn@sdu.dk}
\emailAdd{hdel@sdu.dk}

\abstract{
New physics beyond General Relativity impacts black-hole spacetimes. The effects of new physics can be investigated in a largely theory-agnostic way by following the principled-parameterized approach. In this approach, a classical black-hole metric is upgraded by following a set of principles, such as regularity, i.e., the absence of curvature singularities. We expect these principles to hold in many theories beyond General Relativity.
In the present paper, we implement this approach for time-dependent spacetimes describing gravitational collapse. We find that the Vaidya spacetime becomes regular through the same modification of the spacetime metric as stationary black-hole spacetimes [1-3]. We investigate null geodesics and find indications that the modification is even sufficient to render null geodesics future complete. Finally, we find that the modification of the spacetime structure results in violations of the null energy condition in a finite region inside the apparent horizon of the black hole that forms. Null geodesics are attracted to the boundary of this region, such that the new-physics effects are shielded from asymptotic observers. An exception occurs, if the classical spacetime has a naked singularity. Then, the upgraded spacetime is singularity-free and null geodesics from the regular core can escape towards asymptotic observers.
}

\begin{document}
\maketitle
\flushbottom
\allowdisplaybreaks

\section{Introduction: gravitational collapse and physics beyond GR}
The true nature of black holes is a persistent riddle with many facets. The physics that resolves curvature singularities, makes black-hole spacetimes geodesically complete and removes artifacts such as closed timelike curves and Cauchy horizons (as they occur in the interior of the Kerr spacetime) is unknown, but generally expected to be quantum in nature. 

Investigating black-hole spacetimes beyond General Relativity (GR) can be done in three distinct ways: (i) in a particular (quantum) theory of gravity, (ii) in a general parameterization of the spacetime metric, (iii) in a combination of (i) and (ii), in particular, within the \emph{principled-parameterized} approach to black-hole spacetimes beyond GR, introduced in \cite{Eichhorn:2021etc,Eichhorn:2021iwq,Eichhorn:2022oma}. In this approach, principles from theories beyond GR are used to motivate properties of families of spacetime metrics. This brings together advantages from approaches (i) and (ii), namely the direct connection between theoretical principles and spacetime properties of (i) and the comprehensiveness and generality of (ii).  The principled-parameterized approach is thus largely theory agnostic in that it implements principles that are expected to hold in more than just a single theory beyond GR.

In the present paper, we extend the development of the principled-parameterized approach to spacetimes which are not stationary and which describe gravitational collapse to a black hole.

Metrics arising in the principled-parameterized approach can also be motivated from other settings. For instance, stationary black-hole metrics motivated from Asymptotic Safety, reviewed in \cite{Platania:2023srt}, have been shown to respect all principles, see \cite{Eichhorn:2022bgu} for an explicit discussion of the connection.

Finally, simple models of gravitational collapse are of interest because they can give rise to naked singularities and thus violate the cosmic censorship conjecture, see \cite{Joshi:2011rlc} for a review.\footnote{Those examples may not be astrophysically realized, because they correspond to spherically symmetric  
collapse.} We are interested in exploring whether the singularity can be resolved without the introduction of a horizon, such that the region in which spacetime modifications (e.g., due to quantum gravity) are large, is visible to external observers. 

This paper is organized as follows: in Sec.~\ref{sec:Classical_part}, we review the main properties of the classical Vaidya spacetime which is a simple model of spherically symmetric gravitational collapse. 
In Sec.~\ref{sec:Quantum_part}, we consider an upgrade of the classical Vaidya spacetime in the principled-parameterized approach as well as in an asymptotic-safety inspired setting. 
We provide an outlook in Sec.~\ref{sec:Outlook} and conclude in Sec.~\ref{sec:Conclusion}. Additional technical details can be found in the appendix.

\section{
Review of the classical Vaidya spacetime
}
\label{sec:Classical_part}
To propose singularity-free models of gravitational collapse, we start from the classical Vaidya spacetime. To set the stage for our analysis, we review the line-element and salient features of the spacetime, including the behavior of curvature invariants, the various notions of horizons and the motion of null geodesics near the center.

\subsection{Classical Vaidya spacetime metric
}
The line-element of the classical advanced Vaidya metric 
in advanced Eddington-Finkelstein (EF) coordinates $(v,r,\theta, \phi)$ is\footnote{Throughout this paper we are using units in which $c = \hbar = 1$, but keep $G_0$, for reasons that become clear when we develop a model based on asymptotically safe quantum gravity.}
\be
ds^2_{\rm EF} = - f[v] dv^2 + 2 dv\, dr+ r^2 d\Omega^2,\quad f[v] = 1-\frac{2G_0 m[v]}{r}.
 \label{eq:Vaidyametric}
\ee
To encode the accretion of null dust, the time-varying Misner-Sharp mass function satisfies $m[v] \geq 0$. In the limiting case $m[v]= \rm const$, the line-element describes Schwarzschild spacetime. 
\\

We  focus on the Vaidya-Kuroda-Papapetrou (VKP) model \cite{Vaidya:1966zza, kuroda1984naked, papapetrou1985formation}, in which accretion happens during a finite time. Thus, the mass function is a piecewise function with a phase of linear growth:
\be
m[v] = 
    \begin{dcases}
        0, & v \leq 0 \\
        \mu\, v & 0 < v < \bar{v} \\
        m = \mu\, \bar{v} & v \geq \bar{v}.
    \end{dcases}   
\label{eq:VKPmass} 
\ee
At time $v \leq 0$, the spacetime is locally isometric to Minkowski spacetime. At $v = 0$, ingoing shells of null dust start to collapse under their own gravity. The amount of matter falling towards the center at $r=0$ is encoded in the accretion rate $\mu$. Astrophysically realistic accretion rates are estimated as $G_0 \mu \leq 10^{-8}$, see \cite{Solanki:2022glc}. We consider larger values of $G_0 \mu$ throughout, in order to simplify our numerical studies. 
Theoretically, a critical value of $\mu$ lies at $\mu_c=\frac{1}{16 G_0}$, because a singularity forms in the spacetime before a horizon forms. Thus, the VKP model is one of the first counterexamples \cite{kuroda1984naked} to Penrose's cosmic censorship conjecture \cite{Penrose:1969pc}. In our analysis, we will thus consider both $\mu>\mu_c$ and $\mu<\mu_c$.\\
In all cases, the spacetime finally settles down to a stationary black-hole spacetime, i.e., for $v>\bar{v}$, the spacetime is locally isometric to a Schwarzschild spacetime with ADM mass $m= \mu\, \bar{v}$.

\subsection{Energy conditions}
\label{sec:Classical_energy_conditions}
The spacetime described by Eq.~\eqref{eq:Vaidyametric} contains a curvature singularity at $r=0$ (for $v>0$) and is geodesically incomplete, as we review below. In singularity theorems in GR, one of the necessary assumptions to show geodesic incompleteness is an energy condition for the infalling radiation. We review this condition for the Vaidya spacetime, because we  later examine whether and how it fails in a singularity-free, upgraded spacetime.

To describe the gravitational collapse of ingoing null dust, the energy-momentum tensor constructed from the four-velocity $n_{\mu} \equiv \delta^0_{\mu}$ of the null dust must be pressureless.\footnote{It holds that $n_{\mu}n^{\mu}=0$, because for the line-element Eq.~\eqref{eq:Vaidyametric}, $g^{00}=0$.} In terms of 
the energy density 
\be
\rho  \equiv T_{00} = \frac{\dot{m}[v]}{4 \pi r^2},
\label{eq:EnergyDensity} \quad \mbox{where }\dot{m}[v] \equiv \frac{\partial m[v]}{\partial v},
\ee
we can write
\be
T_{\mu \nu} = \rho\, n_{\mu} n_{\nu}. 
\ee
If the energy-momentum tensor satisfies the appropriate pointlike energy condition, the formation of a spacetime singularity which renders the spacetime geodesically incomplete, is inevitable \cite{Penrose:1964wq}. In particular, the weakest pointlike energy condition is the Null Energy Condition (NEC)  which requires
\be
\varepsilon = T_{\mu \nu} k^{\mu} k^{\nu} \geq 0
\label{eq:NEC}
\ee
for every future-pointing null vector field $k_{\mu}$. It restricts the mass function of the Vaidya spacetime
\be
T_{00}\equiv \rho \geq 0 \stackrel{\text{Eq.}~\eqref{eq:EnergyDensity}}{\Longleftrightarrow} \dot{m}[v] \geq 0.
\ee
This ensures that ``well-behaved'' null dust has to undergo
gravitational collapse (as opposed to the opposite process, namely evaporation, for which $\dot{m}[v]<0$).
This condition is satisfied (for all $v$) by the linearly growing mass function in Eq.~\eqref{eq:VKPmass} of the VKP model, for which the NEC simply leads to positive accretion rates $\mu \geq 0$.

\subsection{Spacetime properties}
\label{sec:Classical_spacetime_properties}
The classical Vaidya spacetime is
geodesically incomplete, because it describes gravitational collapse, solves the Einstein equations and satisfies the pointlike null energy condition. In many spacetimes, geodesic incompleteness goes hand in hand with curvature singularities. We will thus review the behavior of null geodesics as well as curvature invariants near $r=0$. Finally, to provide a comprehensive overview of the starting point of our analysis, we will review the various horizons, the photon sphere and the choice of a specific photon surface of the spacetime.

\subsubsection{Null geodesic motion near $r=0$}
It is  sufficient to follow radial null geodesics to show the geodesic incompleteness of the spacetime. They  solve the equation \cite{Mkenyeleye:2014dwa}
\be
\frac{dr}{dv} - \frac{1}{2} \left(1 - \frac{2  G_0m[v]}{r}\right) = 0.
\label{eq:ClassicalImplicitNullGeo}
\ee
This equation can be obtained from the general expression for null geodesics, 
\be
\frac{d^2 x^{\mu}}{d\lambda^2} + \Gamma^{\mu}_{\alpha \beta} \frac{d x^{\alpha}}{d\lambda} \frac{d x^{\beta}}{d\lambda} = 0,
\label{eq:Geodesic}
\ee
with $\lambda$ an affine parameter and $x^{\mu}[\lambda]
= (v[\lambda], r[\lambda], \theta[\lambda], \phi[\lambda])$ the photon's position in Eddington-Finkelstein coordinates. By specializing to radially infalling geodesics, i.e., 
\be
\frac{d}{d\lambda} \theta[\lambda]
= 0= \frac{d}{d\lambda}  \phi[\lambda],
\ee
we obtain two equations, which describe $r[\lambda]$ and $v[\lambda]$,
\bea
&r \frac{d^2 v}{d\lambda^2} + \frac{G_0 m[v]}{r} \left(\frac{dv}{d\lambda}\right)^2 = 0, \quad
\frac{dr}{d\lambda} - \frac{1}{2} \left(\frac{dv}{d\lambda}\right) \left(1-\frac{2 G_0 m[v]}{r}\right) = 0.
\label{eq:ClassicalDefiningNullGeo}
\eea
The separate dependence of $r$ and $v$ on $\lambda$ can be traded for a dependence of $r$ on $v$ and the two equations \eqref{eq:ClassicalDefiningNullGeo} can be combined into Eq.~\eqref{eq:ClassicalImplicitNullGeo}.

For any value of $\mu$, it has been shown that Eq.~\eqref{eq:ClassicalImplicitNullGeo} admits an analytical, implicit general solution \cite{BKP} of the form
\be
-\frac{2 \arctan{\left(\frac{v - 4\,r[v]}{v \sqrt{-1 + 16 G_0 \mu}}\right)}}{\sqrt{-1 + 16 G_0 \mu}} + 2\,\log{[v]} + \log{\left[2\, \mu G_0 - \frac{r[v]}{v} + \frac{2\, r^2[v]}{v^2}\right]} = C,
\label{eq:SolutionClassicalImplicitNullGeo}
\ee
with $C$ an arbitrary complex integration constant. A set of outgoing null geodesics is obtained by varying the constant $C$.

Whereas the above solution Eq.~\eqref{eq:SolutionClassicalImplicitNullGeo} holds for all values of $\mu$, a simpler representation of the solutions exists 
for $\mu \leq \mu_{\rm c}= \frac{1}{16 G_0}$, found in 
\cite{israel1985general,israel1986formation}, 
\be
\frac{\left|r[v] - \lambda_- v\right|^{\,\lambda_-}}{\left|r[v] - \lambda_+ v\right|^{\,\lambda_+}} = \tilde{C},
\label{eq:ClassicalImplicitNullGeo2}
\ee
with $\tilde{C}$ an arbitrary real and positive constant and
\be
\lambda_{\pm} = \frac{1 \pm \sqrt{1 - 16 \mu G_0}}{4}.
\label{eq:LinearSolutionsClassicalDefiningNullGeo}
\ee

\begin{figure}[h!]
\centering
\subfloat{\includegraphics[width=0.3\linewidth]{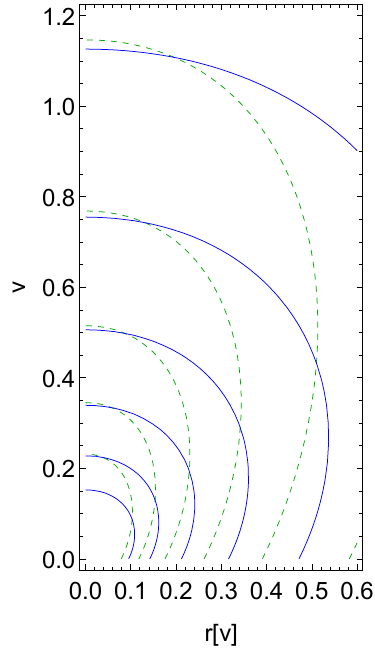}} 
\subfloat{\includegraphics[width=0.305\linewidth]{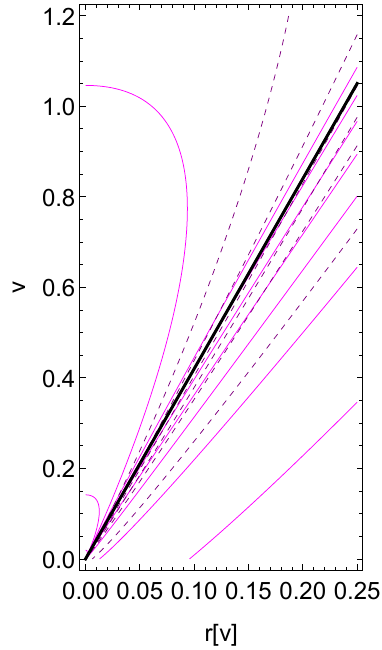}}
\caption{\label{fig:ngs} Left panel: We show null geodesics near $r=0$ for $G_0 \mu=1/2$ (green dashed lines) and $G_0 \mu=1$ (blue continuous lines). Right panel: We show null geodesics near $r=0$ for $G_0 \mu=1/16.5$ (magenta continuous lines) and $G_0 \mu=1/15.5$ (purple dashed lines) and a tangent to a geodesic near the origin $(r = 0, v = 0)$ (black continuous line). The critical value is $G_0 \mu_{\rm c} = 1/16$, as a subcritical geodesic crosses its tangent at the origin, while a supercritical geodesic does not. All plots in this paper are in Planck units in which, in addition to $\hbar = c = 1$, $G_0 = 1$.}
\end{figure}

We now consider  null geodesics near $r=0$ to confirm that the spacetime is geodesically incomplete.
From Fig.~\ref{fig:ngs}, we find the expected result: outgoing null geodesics are  deflected towards $r=0$, which they reach in a finite amount of (advanced and affine) time. The gravitational lensing increases when either $G_0$ or $\mu$ increases, i.e., when gravity becomes stronger or the energy density is increased. This already provides us with a first hint on the two alternatives of how to avoid geodesic  incompleteness in a spacetime describing gravitational collapse, namely altering the effective mass function (or accretion rate) or altering the strength of the Newton coupling. We will explore both alternatives in Sec.~\ref{sec:Quantum_part}.\\ We also see that the behavior of null geodesics changes around $G_0 \mu_{\rm c} = 1/16$, below which null geodesics starting out at $(r,v)=(0,0)$ can actually escape to infinity.  This results in a naked singularity, which 
we will discuss  in more detail below.

\subsubsection{ 
Curvature invariants
}
\label{subsubsec:BehaviourClassicalInvariants}
Geodesic incompleteness and singular curvature invariants are two independent concepts. However, in many black-hole spacetimes, the two go hand in hand, such that an infalling observer will experience both a finite future and diverging tidal forces. Here, we confirm the singular behavior of curvature invariants near $r=0$.

Under rather generic assumptions (spelled out in \cite{Coley:2009eb}), a spacetime metric can be characterized by an algebraically complete basis of seventeen  non-derivative curvature invariants, referred to as Zakhary-McIntosh (ZM) invariants \cite{Carminati:1991ddy, Zakhary:1997xas, carminati2002algebraic}, built out of the Weyl tensor $C_{\mu \nu \rho \sigma}$, the (left-)dual Weyl tensor $\overline{C}_{\mu\nu\rho\sigma} = \frac{1}{2} \epsilon_{\mu\nu \kappa\lambda }C^{\kappa\lambda }_{\phantom{\kappa\lambda }\rho\sigma}$ (with $\epsilon_{\mu\nu \kappa\lambda}$ the totally anti-symmetric Levi-Civita tensor) and the Ricci tensor $R_{\mu \nu}$. The set of invariants decomposes into four real Weyl-invariants $I_{1-4}$, four real Ricci-invariants $I_{5-8}$ and nine real mixed invariants $I_{9-17}$ as listed in App.~\ref{app:ZMinvariants}.\\

For spacetimes which admit Killing vectors, one typically finds that not all of the non-zero invariants are independent. For the classical Vaidya spacetime with VKP mass function, the only non-zero, non-derivative curvature invariants are
\be
I_1 = C_{\mu \nu \rho \sigma} C^{\mu \nu \rho \sigma} = \frac{48 G_0^2 \mu^2 v^2}{r^6},\quad I_3 = C_{\mu \nu}^{\rho \sigma} C_{\rho \sigma}^{\alpha \beta} C_{\alpha \beta}^{\mu \nu} = \frac{96 G_0^3 \mu^3 v^3}{r^9} = \frac{1}{2 \sqrt{3}}\, I_1^{3/2}
\label{eq:ZMinvariantsVKP}
\ee
for $v < \bar{v}$. Because the third invariant $I_3$ can be expressed in terms of $I_1$, $I_1$ is the only independent, non-zero, non-derivative curvature invariant. $I_1$ is clearly singular at the center $r = 0$ for all times $v$.

\subsubsection{Apparent, event and Cauchy horizons}
Horizons in a Vaidya spacetime are discussed in detail in \cite{Griffiths:2009dfa,nielsen2014revisiting}. Here we only briefly review their nature and their location for the particular example of the VKP model.

Stationary black-hole spacetimes are usually characterized by their event horizon. To locate the event horizon, global knowledge of the entire spacetime is required, because the event horizon is the boundary of the causal past of future null infinity. In a time-dependent spacetime, another notion of horizon is often more useful, namely an apparent horizon, reviewed in \cite{Ashtekar:2004cn,Booth:2005qc,Gourgoulhon:2008pu}.
At an apparent horizon, the expansion of both in- and out-going null geodesics is negative semi-definite, i.e., gravitational lensing is so strong that locally, all geodesics are prevented from reaching larger distances from the center. 

For the VKP model, this condition translates into  \cite{Nielsen:2005af, Faraoni:2013aba} 
\be
g^{rr}_{\rm EF} = g_{\rm vv,\, EF} = 0 
\ee
We rederive this in App.~\ref{app:DefEqApparentHorizon} in a form that can be generalized when we modify the spacetime. For the VKP line element \eqref{eq:Vaidyametric}, this condition can be solved to obtain
\be
g_{\rm vv,\, EF} = 0 \Leftrightarrow 1- \frac{2G_0 \mu v}{r_{\rm AH}} = 0 \Leftrightarrow r_{\rm AH} = 2 G_0 \mu v.
\label{eq:ClassApparentHorizon}
\ee

To find the event horizon for the VKP model, we use the initial condition $r_{\rm EH}[v = \bar{v}] = r_{\rm EH,\, Schw} = 2 G_0\, \mu \, \bar{v}$ for the null geodesic equation \eqref{eq:ClassicalImplicitNullGeo} and 
 numerically  solving the equation backward in time to find the event horizon at $v < \bar{v}$.

For $\mu<\mu_{\rm c} = \frac{1}{16 G_0}$, a third notion of horizon is realized, namely a Cauchy horizon. A Cauchy horizon exists when the initial-value problem is no longer well-defined, i.e., when initial data defined on a spatial hypersurface is not sufficient to determine  their future evolution. The Cauchy horizon delineates the boundary of the  spacetime region in which the future evolution of the initial data is not well-defined. Thus, Cauchy horizons appear in particular when there are naked singularities.
In Fig.~\ref{fig:ngs}, we already saw a hint that there is a naked singularity for $\mu<\mu_{\rm c}$, because the right panel shows some null geodesics emanating from $(r,v)=(0,0)$ which are not focused back towards small $r$. The singularity in the VKP model has both a spacelike part and a lightlike part, a section of the latter part is not hidden behind the horizon. It is only from this naked lightlike section that null geodesics starting at $(r,v)=(0,0)$ can reach an asymptotic observer.

The Cauchy horizon can be found from Eq.~\eqref{eq:ClassicalImplicitNullGeo2}, because it is itself a null surface. Eq.~\eqref{eq:ClassicalImplicitNullGeo2} admits two linear solutions, $r_{\pm}[v] = \lambda_{\pm} \cdot v$, which both emanate from the point $(r,v)=(0,0)$. It turns out that $r_-$ is the tangent to the event horizon at that point, cf.~Fig.~\ref{subfig:PhaseDiagClassNullGeoMuSmall}. 
Therefore, all null geodesics that emanate from the point $(r,v)=(0,0)$ and are not focused back towards $r=0$ in the future lie within the wedge in between $r_-$ and $r_+$ and this wedge is non-empty. Accordingly, $r_+$ constitutes the Cauchy horizon of the spacetime.

\subsection{Photon sphere and choice of a specific photon surface}
	\label{sec:Photon sphere}
To complete our discussion of the VKP model and null geodesic motion within it, we review what is known about the photon sphere and photon surfaces and we motivate the choice of a specific photon surface.

Because the VKP model is time-dependent, there is no Killing vector associated to stationarity. However, there is a conformal Killing vector, see \cite{nielsen2014revisiting}, which renders the null geodesic motion separable for all values of $\mu$. Only for $\mu < \mu_{\rm c}$ is the conformal Killing vector timelike in the wedge delineated by the two conformal Killing horizons $r_{\pm}[v] = \lambda_{\pm} v$, which can be used to find the location of the (unique) photon sphere in that wedge.\footnote{We define the photon sphere according to the nomenclature of \cite{Claudel:2000yi}, which corresponds to the most commonly used definition. However, note that this definition differs from that in \cite{Mishra:2019trb}.} The location of the photon sphere at $r_{\rm ps}[v] = \frac{v}{2} \left(1 - \sqrt{1 - 12 G_0 \mu}\right)$ was obtained in \cite{Solanki:2022glc}. The photon sphere is such that any null geodesic initially tangent to its surface will remain tangent to it and its radial location is fixed to $r_{\rm ps} = {\rm const}$. Additionally, the photon sphere gives rise to strong gravitational lensing of lightlike geodesics as past-oriented null geodesics asymptotically approach the photon sphere by spiralling towards it, thus completing one or more half-orbit(s) around the centre $r = 0$.

Our main goal is to analyze the upgraded spacetimes that will be constructed in Sec.~\ref{sec:Quantum_part} and their differences to the classical Vaidya spacetime. As we will discuss, the conformal Killing vector field is lost in the upgraded spacetimes. Thus the construction of a photon sphere as done in \cite{Solanki:2022glc} is not possible. We instead focus on surfaces which exist both in the Vaidya spacetime and in the upgraded spacetimes for all values of $\mu$, namely \emph{photon surfaces} (dubbed \emph{$SO(3)$-invariant photon surfaces} in the nomenclature of \cite{Claudel:2000yi}). In a spherically symmetric spacetime, there are infinitely many photon surfaces and a given individual photon surface is generated by applying all the $SO(3)$ transformations to a given null geodesic. As for the photon sphere, null geodesics initially tangent to a given photon surface will remain tangent to it, but the radial location of a given photon surface can evolve in time. Here we focus on a specific photon surface, hereafter named \emph{the distinguished photon surface}, which is the unique time-evolving photon surface $r_{\rm p}[v]$ at $0 \leq v < \bar{v}$ which matches onto the Schwarzschild photon sphere at $v > \bar{v}$. We obtain the distinguished photon surface by solving the null geodesic equation Eq.~\eqref{eq:VKPEquationPhotonSphere} and supplementing it with the appropriate initial conditions at $v > \bar{v}$ that define the Schwarzschild photon sphere.\\

\begin{figure}[h!]
\centering
\includegraphics[width=0.7\linewidth]{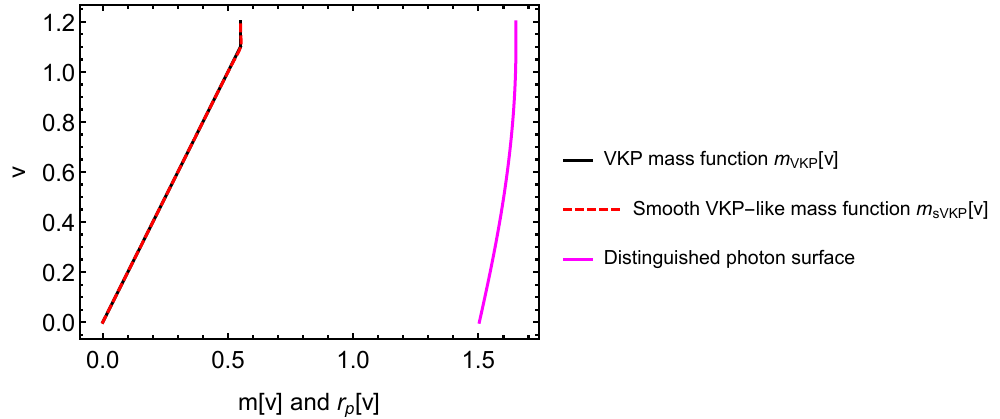}
\caption{Graphical representation of the VKP mass function (black; see Eq.~\eqref{eq:VKPmass}), its smooth approximation with $k=30$ (dashed red; see Eq.~\eqref{eq:SmoothApproxVKPmass}) and the evolution of the distinguished photon surface for the latter mass function with $G_0 \mu = \frac{1}{2}$ (magenta). The accretion stops at $\bar{v} = 1.1$.\label{fig:ClassicalMassesAndPhotonSphereMuVeryLarge}
}
\end{figure}

Due to spherical symmetry, we can restrict ourselves to the equatorial plane  without loss of generality. 
Note that in contrast to Eq.~\eqref{eq:ClassicalImplicitNullGeo}, we are not restricting to radial geodesics here and refer the reader to \cite{Mishra:2019trb} on how to derive the null geodesic equation.
For the VKP model, the null geodesic equation reads
\be
\ddot{r}_{\rm p}[v] - \frac{6 G_0^2 v^2 \mu^2}{r^3_{\rm p}[v]} + \frac{G_0 v \mu (5 - 9 \dot{r}_{\rm p}[v])}{r^2_{\rm p}[v]} + \frac{3 \dot{r}_{\rm p}[v] - 1 + G_0 \mu - 2 \dot{r}^2_{\rm p}[v]}{r_{\rm p}[v]} = 0.
\label{eq:VKPEquationPhotonSphere}
\ee
We then impose initial conditions at $v > \bar{v}$ that correspond to the photon sphere in the Schwarzschild spacetime
\be
\left\{ 
\begin{aligned}
&r_{\rm p}[v] = 3 G_0 \mu \bar{v}, \\
&\dot{r}_{\rm p}[v] = 0,
\end{aligned}
\right.
\label{eq:FutureClassBoundaryConds}
\ee
and follow back the unique solution $r_{\rm p}[v]$ defining the distinguished photon surface. Numerically, we find that the sharp transition occurring in $m[v]$ at $v = \bar{v}$  impacts the numerical stability of the evolution equation for the distinguished photon surface. Hence, in practice, we derive the location of the distinguished photon surface for a smooth function that well approximates the VKP mass, i.e.,
\be 
m_{\rm sVKP}[v] = \frac{\mu v}{1 + e^{-2k(\bar{v} - v)}} + \frac{\mu \bar{v}}{1 + e^{-2k (v - \bar{v})}},
\label{eq:SmoothApproxVKPmass}
\ee
with $k$ sufficiently large. Fig.~\ref{fig:ClassicalMassesAndPhotonSphereMuVeryLarge} shows the original VKP mass function, its smooth version corresponding to Eq.~\eqref{eq:SmoothApproxVKPmass} 
and the distinguished photon surface for $G_0 \mu = \frac{1}{2}$.

The photon sphere marks the location of strong gravitational lensing of null trajectories. For static black holes, the photon sphere surrounds the event horizon at a radius not much larger than the radius of the event horizon (see the spacetime diagrams in Fig.~\ref{fig:PhaseDiagClassNullGeo} for $v > \bar{v} = 1.1$). However, in a dynamical spacetime like the VKP model, the photon sphere does not exist for all values of $\mu$, while the distinguished photon surface does and coincides with Schwarzschild's photon sphere once the accretion stops. The distinguished photon surface follows the infalling shells of null dust. For very low accretion rates, the distinguished photon surface starts at finite $v > 0$ at a surface $v = {\rm const.}$ located far away from the center, c.f.~Fig.~\ref{subfig:PhaseDiagClassNullGeoMuSmall}, see also \cite{Claudel:2000yi}. In contrast, for very large accretion rates $1 > G_0 \mu \gg G_0 \mu_{\rm c}$, shells of null dust quickly fall towards the center, such that the distinguished photon surface forms at early times $v \gtrsim 0$ and grows from a finite radius close to the radius of Schwarzschild's photon sphere, see Fig.~\ref{subfig:PhaseDiagClassNullGeoMuVeryBig} and \cite{Koga:2022dsu}. Finally, the case $\mu \gtrsim \mu_{\rm c}$ is in between these two: the distinguished photon surface starts at large radii as for very small accretion rates, but at earlier times $v \gtrsim 0$, moves inwards as it follows the accreting null dust and later moves towards the radius of Schwarzschild's photon sphere.

\subsubsection{Spacetime diagrams of the VKP model}
\label{sec:Classical_phase_diagram}

\begin{figure}[!t]
\centering
\subfloat[\label{subfig:PhaseDiagClassNullGeoMuVeryBig}
$1 > G_0\mu >>  G_0 \mu_{\rm c}$.]{\includegraphics[width=0.45\linewidth]{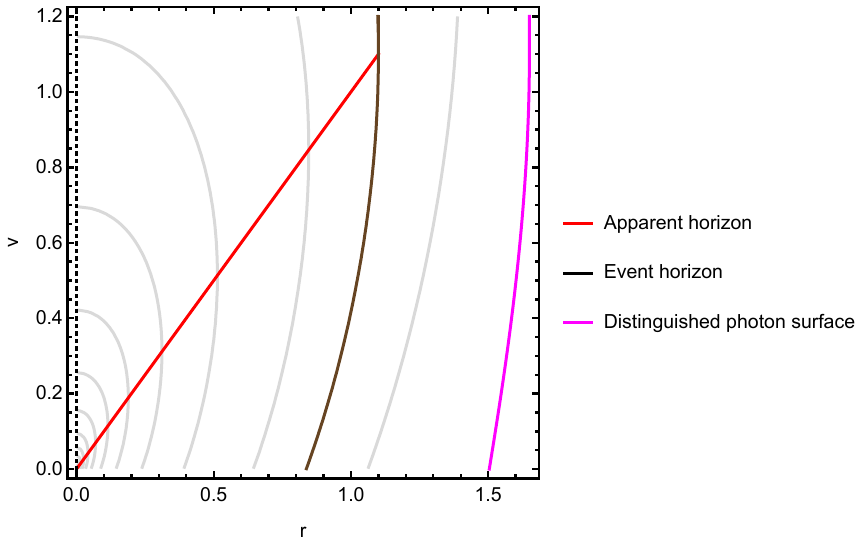}}
\subfloat[\label{subfig:PhaseDiagClassNullGeoMuBig}
$G_0 \mu \gtrsim G_0\mu_{\rm c}$.]{\includegraphics[width=0.45\linewidth]{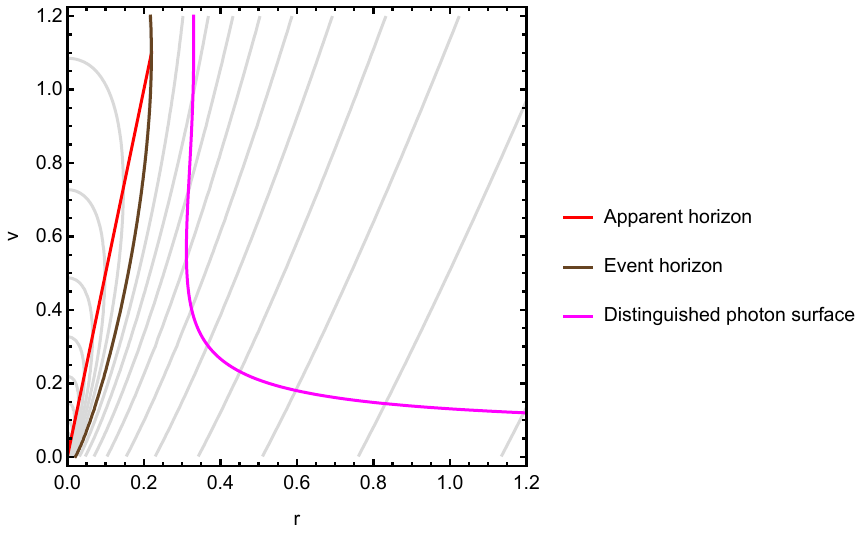}}
\hspace{\fill}
\subfloat[\label{subfig:PhaseDiagClassNullGeoMuSmall}
$\mu \leq \mu_{\rm c}$.]{\includegraphics[width=0.45\linewidth]{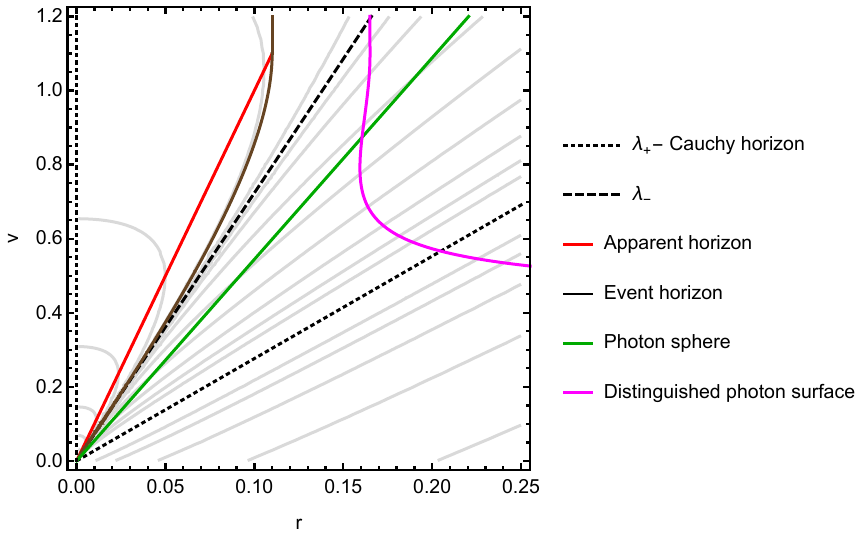}}
\caption{Spacetime diagrams $(r[v], v)$ of null geodesics for the classical VKP model. Top left panel: for $G_0 \mu = \frac{1}{2} \gg
G_0 \mu_{\rm c}$ the curvature singularity is hidden behind an event horizon. Top right panel: $G_0 \mu = \frac{1}{10} > G_0 \mu_{\rm c}$ (similar behavior). Bottom panel: 
for
$G_0 \mu = \frac{1}{20} < G_0 \mu_{\rm c}$ a globally naked singularity is present and the distinguished photon surface intersects the Cauchy horizon $r_+[v]$ \cite{Claudel:2000yi} and the photon sphere (in green) \cite{Solanki:2022glc} before reaching the location of the Schwarzschild photon sphere. The accretion stops at $\bar{v} = 1.1$  for all three cases.\label{fig:PhaseDiagClassNullGeo}}
\end{figure}

We can now summarize our review of the classical VKP model in the three spacetime diagrams in Fig.~\ref{fig:PhaseDiagClassNullGeo}. 
The spacetime diagrams show two distinct regimes, depending on the value of the accretion rate $\mu$ compared to its critical value $\mu_{\rm c} \equiv \frac{1}{16 G_0}$. In both regimes, the classical VKP model is geodesically incomplete, as shown by the finite advanced time $v$ taken by light rays to fall back to the singularity located at the center $r = 0$.\footnote{While future null geodesic incompleteness necessitates that outgoing null geodesics reach the central singularity at finite advanced time $v$, this is not sufficient. It is a sufficient condition that $r[\lambda]$ reaches the center at $r = 0$ and $v[\lambda]$ becomes constant at a finite value of the affine parameter $\lambda$. The behavior of null geodesics in a Vaidya spacetime has been extensively studied, e.g.~in \cite{Griffiths:2009dfa,Joshi:93glo} which confirm geodesic incompleteness.} This is consistent with the NEC condition  and the Einstein equations holding, cf.~Sec.~\ref{sec:Classical_energy_conditions},  as implied by the null version of the Penrose-Hawking singularity theorems. 
 
The spacetime always possesses an apparent horizon given in Eq.~\eqref{eq:ClassApparentHorizon} which forms a straight line contained inside an event horizon, both forming at $v = 0$. 

The curvature singularity in the center $r = 0$ is always covered by an event horizon for $\mu > \mu_{\rm c}$, i.e., for all $v \geq 0$, satisfying the strong cosmic censorship conjecture. However, for $\mu \leq \mu_{\rm c}$, shown in Fig.~\ref{subfig:PhaseDiagClassNullGeoMuSmall}, a globally naked singularity forms: all null rays comprised between the two Killing horizons $r_{\pm}[v]$ emanate from the singularity at $(r[v], v) = (0,0)$ and can reach an observer at infinity. Thus the spacetime contains a Cauchy horizon that forms the causal boundary of the region in which the initial-value problem has a well-defined future evolution.

Hence, despite its simplicity, the VKP model for spherically symmetric gravitational collapse can violate the strong cosmic censorship conjecture in the regime $\mu \leq \mu_{\rm c}$.

\section{
Regular gravitational collapse in two approaches
}
\label{sec:Quantum_part}
We put forward a family of upgraded Vaidya metrics that we construct by implementing a set of principles for phenomenological models of spacetimes beyond GR.  Our starting point is the VKP model, and the upgraded spacetimes are all generalized Vaidya spacetimes in which $m= m[v,r]$.
Incidentally, one member of this family of metrics can be constructed in a different, independent approach, namely by Renormalization-Group (RG) improvement within asymptotically safe gravity. 

We start this section by reviewing the principled-parameterized approach to stationary spacetimes beyond GR as well as RG improvement in asymptotically safe gravity and explaining what the relation between the two approaches is.

The principled-parameterized approach, developed in \cite{Eichhorn:2021iwq, Eichhorn:2021etc,Eichhorn:2022oma} is largely agnostic with respect to the theory of gravity. It is a phenomenological approach, in which the guiding question is: what is the minimal modification of a given singular, classical spacetime that implements four principles, namely  locality, simplicity, regularity and a Newtonian limit? In practice, for stationary black-hole spacetimes, the minimal modification consists in an upgrade of the ADM mass parameter to a function of the spacetime coordinates. Based on \cite{Eichhorn:2021iwq, Eichhorn:2021etc,Eichhorn:2022oma}, this mass modification is sufficient to implement all principles; based on \cite{Delaporte:2022acp}, an upgrade of the spin parameter is not. To implement locality, the ADM mass is not upgraded to a general function of the spacetime coordinates, but instead depends on a coordinate-invariant quantity, namely a suitable choice of local curvature invariants. Because the upgraded mass function must have a dimensionless argument, the upgrade introduces a new scale into the spacetime, namely the new-physics scale $r_{\rm NP}$. Simplicity is achieved if no second scale is introduced. The last two principles, regularity (i.e., absence of curvature singularities) and the Newtonian limit dictate the asymptotic dependence of the mass function on the curvature; simplicity dictates monotonicity of the mass function in between the two asymptotic limits. Thereby, the four principles are sufficient to arrive at a family of  regular metrics. This family has one free function $m_{\rm NP}[x^{\mu}]$, of which the asymptotic behaviors are fixed, and one free scale $r_{\rm NP}$ that determines the transition between the two asymptotic behaviors.

Instead, RG improvement starts from a specific theory, typically asymptotically safe gravity, and is thus part of what one might call a ``principled'' approach to spacetimes beyond GR. RG improvement is a method to incorporate loop corrections into the solutions of the equations of motion and is as such well-established in quantum field theory \cite{Coleman:1973jx}. For asymptotically safe gravity, it has first been used in \cite{Bonanno:1998ye,Bonanno:2000ep}. However, in the context of gravity, there are ambiguities  in the procedure, see, e.g., \cite{Held:2021vwd}, reviewed in \cite{Eichhorn:2022bgu}. Thus, the resulting black-hole spacetimes have the status of toy models inspired by asymptotically safe gravity, rather than solutions to a full theory of quantum gravity.\\
In its simplest incarnation, RG improvement starts from a classical spacetime and promotes the coupling constants to scale dependent couplings that ``run'' as a function of scale, as described by the RG. The final step consists in identifying the RG scale with a suitable physical scale of the spacetime. This step is well-motivated in asymptotically safe gravity, in which the specific version of an RG equation that is used is the functional RG, see \cite{Dupuis:2020fhh} for a review. The functional RG is based on an infrared cutoff in the path integral. This infrared cutoff is lowered successively, such that fluctuations are integrated over step by step. The decoupling mechanism causes fluctuations to decouple once their mass scale is reached, i.e., fluctuations in a given field no longer impact the effective dynamics once the infrared cutoff lies below their mass scale. Because curvature can act as an effective mass, the identification of the RG scale with a curvature scale is well-motivated, see \cite{Platania:2023srt, Borissova:2022mgd} for further discussion.\footnote{In the literature, there are examples in which the RG scale is instead equated to an inverse length scale, e.g., the geodesic distance from the black hole's center, e.g., \cite{Bonanno:2000ep,Reuter:2010xb,Falls:2010he,Torres:2014gta}. This is not motivated by the decoupling mechanism, because the geodesic distance does not enter the effective mass of modes. In settings with a high degree of symmetry, e.g., in spherically symmetric black holes, the results are equivalent with those obtained by using the curvature scale as the RG scale. In settings with fewer Killing vectors, the results are no longer equivalent \cite{Eichhorn:2022bgu}.} Taking as our starting point the Vaidya metric, the only coupling in that metric is the Newton coupling $G_0$, which is first promoted to a running coupling $G$ and then chosen to depend on the curvature. 

Incidentally, RG improvement of static black-hole spacetimes results in a metric that also arises in the principled-parameterized approach, see \cite{Eichhorn:2022bgu}.
Going from static to stationary spacetimes, the situation already becomes more subtle, because the locality principle is not always respected when RG improvement is implemented by choosing a non-local notion of scale \cite{Reuter:2010xb,Litim:2013gga,Pawlowski:2018swz}, see the discussion in \cite{Eichhorn:2022bgu}. This results in important differences; e.g., (non-)circularity of the spacetime \cite{Delaporte:2022acp}. Similarly, in time-dependent settings describing gravitational collapse, the regularity principle is not obeyed in examples in which the locality principle is neglected \cite{Bonanno:2016dyv,Borissova:2022mgd}. Here, we aim at exploring whether or not implementing the locality principle gives rise to a regular spacetime.

In fact, we discover that the RG improved Vaidya metric is equivalent to one member in the family of regular metrics constructed in the principled-parameterized approach. The reason is twofold: first, the Newton coupling always appears as a multiplicative factor in front of the mass function in the classical metric, thus an upgrade of the Newton constant to a curvature-dependent function can be traded for an upgrade of the Misner-Sharp mass to a curvature-dependent function. Second, the RG dependence of the Newton coupling satisfies two of the other principles of the principled-parameterized approach automatically. The first principle is simplicity. It holds, because the only special scale in the running of the Newton coupling is the Planck scale, where the transition between the asymptotically safe scaling regime and the classical regime occurs. The second principle  is the Newtonian limit which is realized because in the low-curvature regime, the Newton coupling is constant to recover classical GR from asymptotic safety, thus the metric is not modified in this regime. Non-trivially, the fixed-point scaling of the Newton coupling is also just sufficient to make curvature invariants regular; thus the metric also satisfies regularity and is therefore one special choice in the family of regular metrics.

\subsection{A regular upgrade of the Vaidya metric in the principled-parameterized approach
}
\subsubsection{Implementing the locality principle}
We start from the Vaidya metric with the line element \eqref{eq:Vaidyametric}. In contrast to previous works in the principled-parameterized approach \cite{Eichhorn:2021iwq, Eichhorn:2021etc,Eichhorn:2022oma}, the spacetime is not stationary and the mass not a constant parameter, but already depends on $v$. We promote it to a more general function of the coordinates
\be
m[v] \rightarrow m_{\rm NP}[x^{\mu}],
\ee
where the subscript $\rm NP$ stands for New Physics and indicates that this upgrade should be understood as a phenomenological approach to a more complete theory of gravity beyond GR. To implement locality, $m_{\rm NP}[x^{\mu}]$ may only depend on the spacetime coordinates through curvature invariants.\footnote{Because we work at the level of the spacetime metric and not the action or the equations of motion, we do not know whether locality in the sense of the locality principle that we implement here translates into a local action or not.} This follows an effective-field-theory reasoning, in which higher-order curvature invariants are expected to be present in the action and to modify the metric at high enough values of the classical curvature. 

As discussed in Sec.~\ref{subsubsec:BehaviourClassicalInvariants} above, the only independent non-derivative curvature invariant of the Vaidya spacetime is $I_1$ and thus we choose
\be
m_{\rm NP}[x^{\mu}] = m_{\rm NP}[I_1\, r_{\rm NP}^4] = m_{\rm NP}\Bigl[\frac{48 G_0^2 m[v]^2}{r^6}r_{\rm NP}^4\Bigr].
\ee
The new-physics scale $r_{\rm NP}$ appears in order to keep the argument of the mass function dimensionless.\footnote{Recall that we work in units in which $c=1$ (but $G_0 \neq 1$.)} The upgraded mass function $m_{\rm NP}$ also depends on the classical mass function $m[v]$; to avoid confusion between the original mass function $m[v]$ and the upgraded mass-function, the latter always carries the subscript $\rm NP$, i.e., $m_{\rm NP}[I_1 r_{\rm NP}^4]$. 

\subsubsection{Implementing the Newtonian limit}
At this stage, $m_{\rm NP}$ is a completely arbitrary function, but we now fix its asymptotic behavior at low values of the curvature. We require that for $I_1 r_{\rm NP}^4 \rightarrow 0$, $m_{\rm NP} \rightarrow m[v]$, such that the classical spacetime (which has the appropriate Newtonian limit for $v> \bar{v}$) is recovered. This is easiest implemented by writing
\be
m_{\rm NP}\Biggl[v, \frac{48 G_0^2 m[v]^2}{r^6}r_{\rm NP}^4\Biggr] = m[v]\, f_{\rm NP}\Biggl[\frac{48 G_0^2 m[v]^2}{r^6}r_{\rm NP}^4\Biggr],
\label{eq:upgradedmassfunction}
\ee
where $f_{\rm NP}$ is a function that parameterizes the modification and of which we now require
\be
f_{\rm NP}\Biggl[\frac{48 G_0^2 m[v]^2}{r^6}r_{\rm NP}^4\Biggr] \,\,\overset{\scriptscriptstyle I_1 r_{\rm NP}^4 \rightarrow 0}{\looongrightarrow}\,\, 1.
\ee
 In principle, one can adjust subleading coefficients of $f_{\rm NP}$ in the low-curvature expansion to specific post-GR corrections, e.g., from the EFT approach to quantum gravity, but we will not do so here and keep the subleading coefficients general.

\subsubsection{Implementing regularity}
The other asymptotic limit, at large curvature, is fixed by the regularity principle, i.e., the absence of curvature singularities. To determine the correct fall-off of $f_{\rm NP}[I_1 r_{\rm NP}^4]$ at large $I_1 r_{\rm NP}^4$, we first need to evaluate the curvature invariants with the upgraded mass function. Because the upgraded mass function is also a function of $r$, not just $v$, the upgraded spacetime is part of the class of generalized Vaidya spacetimes \cite{hughston1971generalized},
\be
ds^2_{\rm EF} = - f[v,r] dv^2 + 2 dv\, dr+ r^2 d\Omega^2,\quad f[v,r] = 1 - \frac{2 G_0 m[v,r]}{r}.
\label{eq:GeneralisedVaidyametric}
\ee
We study its complete set of 17 non-derivative curvature ZM invariants. 
Their explicit dependence on the generalized mass function $m[v,r]$ is given in App.~\ref{app:GeneralisedZMinvariants}. With the exception of $I_1$ and $I_3$, all curvature invariants are polynomial in $m'[v,r] \equiv \frac{\partial m[v,r]}{\partial r}$ and $m''[v,r] \equiv \frac{\partial^2 m[v,r]}{\partial r^2}$. This is consistent with the fact that all curvature invariants except for $I_1$ and $I_3$ vanish for the Schwarzschild spacetime, for which $m[v,r] = m = \rm const$.\\

Because the curvature invariants are generically singular as $r \rightarrow 0$, the mass-function must acquire an  $r$-dependence to lift the singularity. To determine the minimal power of $r$ required, we make a power-law ansatz for the small-$r$-limit of the generalized mass function,
\be
m_{\rm NP}[v,r] = h[v]\, r^{\alpha}.
\label{eq:PowerLawAnsatzMassFunction}
\ee
The exact form of $h[v]$ does not matter for this argument.
All non-vanishing Weyl invariants (which include the Weyl tensor or its dual) of the upgraded spacetime vanish for $\alpha = 2, 3$ (unless $h[v] = \rm const.$, in which case $\alpha = 2$ is still singular). However, all Ricci invariants of the upgraded spacetime are singular for $\alpha < 3$, singling out
\be
\alpha = 3
\label{eq:MinimalPowerAnsatzMassFunction}
\ee
as the critical case, i.e., the minimal power of $r$ required to lift the singularity in all polynomially independent non-derivative curvature invariants. This agrees  with the limiting case of the modified Schwarzschild, static spacetime, reached for $h[v] \rightarrow \rm const.$, in which the curvature singularity of the Schwarzschild spacetime is lifted, if the mass is upgraded to a radially dependent function with leading-order behavior $m_{\rm NP}[r] \sim r^3$, cf.~\cite{Eichhorn:2022oma}.

We now translate the results of this asymptotic analysis into a requirement on $f_{\rm NP}[I_1 r_{\rm NP}^4]$. To achieve $m_{\rm NP}[v,r] \sim r^3$ or higher powers of $r$, we must have
\be
f_{\rm NP} \sim \frac{1}{\left(I_1 r_{\rm NP}^4\right)^{\frac{n}{2}}}, \quad \mbox{for } I_1 r_{\rm NP}^4\rightarrow \infty,
\ee
with $n \geq 1$.
We mostly focus on the case $n=1$, which is the minimal case to make the spacetime regular.

\subsubsection{Implementing simplicity}
\label{subsubsec:simplicity}
Simplicity requires that $f_{\rm NP}[I_1 r_{\rm NP}^4]$ introduces only a single new-physics scale into the spacetime. This translates into two constraints on $f_{\rm NP}$: first, it must not depend explicitly on any other scale than $r_{\rm NP}$; second, it must be a monotonic function of $I_1 r_{\rm NP}^4$, because any additional extremum introduces a second scale besides $r_{\rm NP}$.  Thus, $f_{\rm NP}$ must be a monotonic function of a single argument, namely $I_1 \cdot r_{\rm NP}^4$.

The arguably simplest way to achieve all four principles is to choose
\be
f_{\rm NP}[I_1 r_{\rm NP}^4] = \frac{1}{1+\left(I_1 r_{\rm NP}^4 \right)^{\frac{n}{2}}}, \quad n \geq 1.\label{eq:fNP}
\ee
Other functions with the same asymptotic behavior that fulfill the monotonicity requirement, following, e.g., a Dymnikova-profile \cite{Dymnikova:1992ux}, are part of the same family of upgraded Vaidya spacetimes. 
One notable example that is already contained in Eq.~\eqref{eq:fNP} is Hayward's regular gravitational collapse \cite{Hayward:2005gi}, with the identification $\ell^2 = \sqrt{6}r_{\rm NP}^2$ and for the choice $n=1$.

Due to the monotonicity requirement and the two asymptotic constraints, $f_{\rm NP}$ is everywhere smaller  than or equal to one. Thus, the upgraded mass is, at any given advanced time $v$, smaller  than  or equal to the classical mass
\be
m_{\rm NP}[v, I_1 \cdot r^4_{\rm NP}] = m[v] f_{\rm NP}[I_1 \cdot r^4_{\rm NP}] \leq m[v]\quad  \forall v.
\ee
As a result, the modified collapsing body is more compact than its classical counterpart in the sense that, once a horizon is formed, it is more compact than the classical horizon. This is a generic feature of regular black holes satisfying the conditions spelled out in \cite{Eichhorn:2022oma}, for which the curvature singularity is removed by weakening gravity. The only way to avoid the resulting increase in compactness is to introduce a second scale, such that $m_{\rm NP}$ is larger than its classical counterpart in the spacetime region around the apparent horizon and to violate simplicity.

\begin{figure}[!t]
\begin{center}
\includegraphics[width=0.5\linewidth]{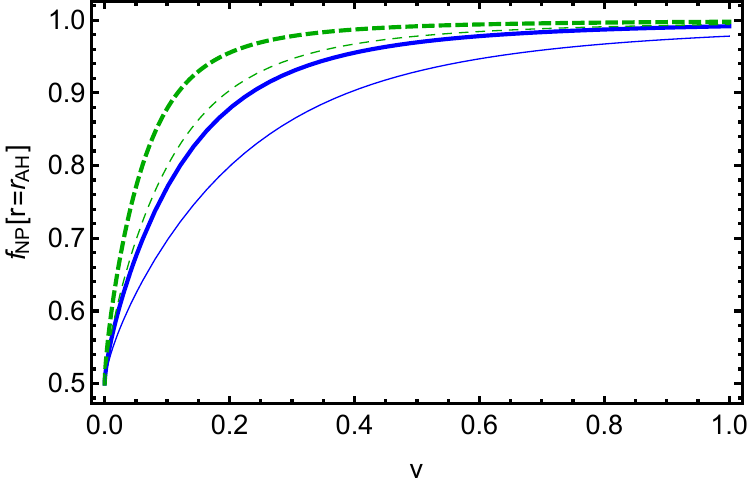}
\end{center}
\caption{\label{fig:fNPAppHorizon} We show $f_{\rm NP}[v, r = r_{\rm AH}]$ evaluated at the location of the apparent horizon $r_{\rm AH}$ Eq.~\eqref{eq:GeneralisedApparentHorizon}, as a function of $v$ for $r_{\rm NP}=10^{-3}$ (thick lines) and $r_{\rm NP}=6\cdot 10^{-3}$ (thin lines). For each value of $r_{\rm NP}$ we consider two accretion rates $G_0 \mu=10^{-2}$ (blue lines) and $G_0 \mu = 2 \cdot 10^{-2}$ (green lines). The classical behavior is recovered in the limit $f_{\rm NP}\rightarrow 1$. Conversely, the more $f_{\rm NP}$ departs from $1$, the stronger are the new-physics effects.}
\end{figure}

To illustrate the increase in compactness, we show $f_{\rm NP}[v,r = r_{\rm AH}]$ as in Eq.~\eqref{eq:fNP} for $n=1$ at the apparent horizon $r_{\rm AH}$ given in Eq.~\eqref{eq:GeneralisedApparentHorizon} in Fig.~\ref{fig:fNPAppHorizon}.

$f_{\rm NP}$ increases as a function of $v$, because the mass of the black hole grows, decreasing the relative impact of new physics at the constant scale $r_{\rm NP}$ over time. Smaller accretion rates result in a smaller $f_{\rm NP}$ at a given advanced time $v$, because the black-hole mass is smaller than for a larger accretion rate. These effects can also be read off from the Taylor expansion of $f_{\rm NP}$ in the dimensionless quantity $\frac{G_0 \mu v}{r_{\rm NP}}$, which is
\be
f_{\rm NP}[r= r_{\rm AH}] = \frac{1}{2} + \frac{\left(G_0 \mu v\right)^{\frac{2}{3}}}{2^{\frac{5}{3}}\cdot 3^{\frac{1}{6}} \cdot r_{\rm NP}^{\frac{2}{3}}} + \mathcal{O}\left( \left(\frac{G_0 \mu v}{r_{\rm NP}}\right)^{\frac{4}{3}} \right).
\label{eq:fNPAtAppHorizon}
\ee

We will analyze further properties of the family of upgraded spacetimes below, but first explain how one member of the family can be constructed inspired by asymptotically safe gravity.

\subsection{A regular upgrade of the Vaidya metric inspired by asymptotically safe gravity
}
\label{subsubsec:AS_and_RG_improved_black_holes}
Asymptotically Safe Quantum Gravity (ASQG) is a quantum field theoretic approach to quantum gravity. Gravity is asymptotically safe, if there is
quantum scale symmetry, encoded in an interacting fixed point of the RG, above the Planck scale. This  ensures UV completion and (non-perturbative) renormalizability in the sense that only a finite number of free parameters must be fixed to make the theory predictive. There is by now compelling evidence for an interacting fixed point in gravity (as well as gravity plus suitable matter) in four dimensions in Euclidean signature, see \cite{Eichhorn:2018yfc,Bonanno:2020bil,Eichhorn:2022gku,Saueressig:2023irs,Pawlowski:2023gym}
for reviews and \cite{Percacci:2017fkn,Reuter:2019byg} for textbooks.
Most of these results rely on the Functional Renormalization Group (FRG) \cite{Wetterich:1992yh,Morris:1993qb}, reviewed in \cite{Dupuis:2020fhh}
and adapted to gravity in the seminal paper \cite{Reuter:1996cp}. For our purposes, it is useful to consider the regularized generating functional on which the method relies, namely
\be
Z_k[J] = \int_{\Lambda_{\rm UV}} \mathcal{D}h_{\mu\nu}e^{-S[\bar{g}_{\mu\nu}+h_{\mu\nu}] - S_{\rm gf}[\bar{g}_{\mu\nu}; h_{\mu\nu}] - \frac{1}{2}\int d^4x\sqrt{\bar{g}}h_{\mu\nu}R^{\mu\nu\kappa\lambda}[-\bar{D}^2/k^2]h_{\kappa\lambda}+ \int d^4x\sqrt{g}J_{\mu\nu}h^{\mu\nu}}.
\ee
In here, the full metric $g_{\mu\nu}$ is split into an auxiliary background $\bar{g}_{\mu\nu}$ and fluctuations $h_{\mu\nu}$. The background is used to gauge-fix the fluctuations through the gauge-fixing term $S_{\rm gf}$ (understood to include the corresponding Faddeev-Popov ghost term). Because the fluctuations are not restricted in amplitude to be small, $Z_k[J]$  is a fully non-perturbative path-integral. $J_{\mu\nu}$ is a source term and $\Lambda_{\rm UV}$ indicates that the path integral has been suitably regularized in the UV. Finally, $R^{\mu\nu\kappa\lambda}[-\bar{D}^2/k^2]$ is an infrared cutoff term which suppresses fluctuations according to their generalized momentum, i.e., the eigenvalues of the background-covariant Laplacian $-\bar{D}^2$: decomposing a field configuration $h_{\mu\nu}[x]$ into eigenmodes of $-\bar{D}^2$, those with eigenvalues higher than $k^2$ are integrated out first. Successively lowering $k$, one integrates out all fluctuations in the path integral. A functional differential equation for the Legendre transform of $Z_k[J]$ tracks the resulting scale-dependence of the couplings.  For instance, the resulting scale dependence of the Newton coupling is, to a good approximation, given by \cite{Reuter:2000nt}
\be
G_N(k) = \frac{G_0}{1+\frac{G_0}{G_{\ast}}k^2},\label{eq:runningG}
\ee
where $G_{\ast}$ is a dimensionless number corresponding to the fixed-point value of the dimensionless product $G_N\cdot k^2$ and we recall that $G_0$ is the classical value of the Newton constant. As a result, for $k^2 \ll \frac{G_0}{G_{\ast}}$, $G_N(k) = G_0 =\rm const$, i.e., at low scales, classical gravity with a constant Newton coupling is recovered. For $k^2 \gg \frac{G_0}{G_{\ast}}$, we enter a scaling regime with $G_N(k) \sim k^{-2}$. This can be understood as a weakening of gravity through gravitational fluctuations. This weakening is a prerequisite for a quantum field theory of gravity to make sense. Among other things, it also suggests that classical curvature singularities could be resolved -- at least, if one associated the behavior of the coupling at high $k^2$ with the behavior at high curvature scales.

This association brings us directly to the idea underlying RG improvement, which is to use the scale dependence \eqref{eq:runningG} in the classical Vaidya metric and set $k^2 = \sqrt{I_1}$. 

Several points should be made clearer at this stage, concerning both the level at which the RG improvement is made and the scale identification. 

Firstly, the upgrade of the classical coupling constants to their running counterparts can be made at three different levels, cf.~the discussion in the reviews \cite{Eichhorn:2022bgu,Platania:2023srt}: in the classical action, in the equations of motion or in the spacetime metric. The three improvements are in general not equivalent, resulting in potentially different upgraded spacetime metrics. It is unclear which actually produces results closest to the solution of the full quantum theory, however, the last one is most straightforward to implement. Secondly, classical gravitational systems usually possess more than one characteristic scale, resulting in some freedom of choice for the scale identification. Even in settings with a high degree of symmetry, where all scales are related to each other, different choices can produce different results. For instance, for spherically symmetric, static black holes, where the ADM mass $m$ sets the value of all other scales, identifying $k$ with, e.g., the local curvature scale \cite{Held:2019xde} or the Hawking temperature \cite{Borissova:2022jqj} produces different results.

With these caveats in mind, we view the spacetime that we construct below as a spacetime \emph{inspired} by asymptotically safe gravity and expect that it may capture some of the features that a solution to the full quantum equations of motion has, but stress that this expectation can only be checked a posteriori.

Adopting the RG-improvement procedure developed for black holes in ASQG, we promote the classical Newton constant $G_0$ to its dimensionful running counterpart in Eq.~\eqref{eq:runningG}
and identify $k^2 \sim \sqrt{I_1}$. 
Incidentally, this corresponds to a particular choice of the upgraded mass function that is included in Eq.~\eqref{eq:upgradedmassfunction}.\\ 

Classical gravitational-collapse spacetimes have been RG-improved within ASQG in \cite{Casadio:2010fw,Torres:2014gta,Torres:2014pea,Torres:2015aga,Bonanno:2016dyv,Bonanno:2017zen,Platania:2019kyx,Borissova:2022mgd,Bonanno:2023rzk}. Different RG-improvement procedures and scale identifications lead to somewhat different conclusions regarding the singularities and the spacetime's structure. However, as a universal result, the central singularity is weakened, as first pointed out in \cite{Casadio:2010fw}. Whether the central curvature singularity is made integrable or fully cured depends both on the type of RG improvement and the choice of scale identification. For example, the central singularity is present but made weaker in \cite{Bonanno:2016dyv,Bonanno:2017zen} when identifying the RG scale with the collapsing fluid's energy density. It is also found to be weakened in \cite{Borissova:2022mgd}, where the scale is determined via the decoupling mechanism \cite{Reuter:2003ca} in an iterative sequence of RG improvements \cite{Platania:2019kyx}. Instead, the central curvature singularity can be fully regularized by the effective repulsive forces generated by the running of the gravitational coupling, as in \cite{Torres:2014gta,Bonanno:2023rzk}. This conclusion still holds when backreaction from Hawking radiation is included, see \cite{Torres:2014pea,Torres:2015aga}. However, the absence of shell-focusing singularities does not entail that shell-crossing singularities are absent at larger radii, as interactions of dust or fluid particles, which would prevent matter shells from crossing, are usually ignored.\\ 
We consider the scale identification with curvature invariants the most physical choice and it is reassuring to see that this choice results in an absence of curvature singularities for static black holes, stationary black holes (as reviewed in \cite{Eichhorn:2022bgu}) and, as we have shown here, for at least one example of gravitational collapse.

\subsection{Null geodesics near $r=0$}
\label{subsec:quantnullgeodesicsnearcentre}
As is well established, regularity of curvature invariants and geodesic completeness are not contingent upon  one another. It is therefore not guaranteed that the spacetime we have constructed by requiring regular curvature invariants is geodesically complete.

\begin{figure}[h!]
\begin{center}
\includegraphics[width=0.4\linewidth]{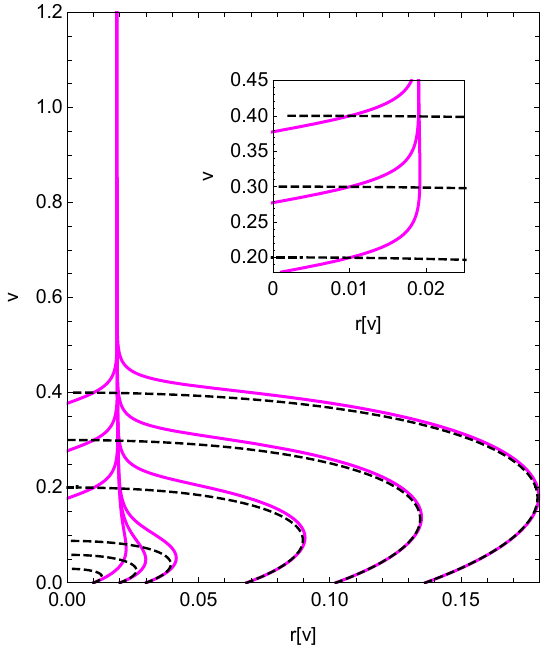}
\end{center}
\caption{\label{fig:nullgeonearrzeroreverse} We show null geodesics near $r=0$ for the upgraded case (magenta, continuous lines) and the corresponding classical case (black dashed lines) for $G_0 \mu=1/2$ and setting $G_0=1$ for the purposes of the plot. We choose $r_{\rm NP}=10^{-2}$. The inset zooms in on a set of trajectories at finite $v$, but very close to $r=0$. The derivative $\frac{dr}{dv}$ has opposite sign in the upgraded case to what it has in the classical case.}
\end{figure}

The radial null geodesic equation for the upgraded spacetime reads
\be
\frac{dr}{dv} = \frac{1}{2}\left(1- \frac{2m[v,r]}{r} \right).
\ee
We consider its small-$r$-behavior, where we can by construction write $m[v,r] = h[v]r^{\alpha}$, with $\alpha \geq 3$ and $h[v]$ a positive function. We thus obtain
\be
\frac{dr}{dv} = \frac{1}{2}\left(1- 2h[v]r^{\alpha-1} \right) \approx \frac{1}{2},\quad \mbox{for } r \rightarrow 0.
\ee
Instead of a divergent and negative right-hand side, as in the classical case, the right-hand side is positive and finite. Accordingly, null geodesics are repelled from the core at $r=0$ and instead move towards larger $r$. We confirm this behavior numerically, cf.~Fig.~\ref{fig:nullgeonearrzeroreverse}.

The change of sign of the derivative $\frac{dr}{dv}$ from negative in the classical case to positive in the upgraded case suggests that geodesics are no longer future-incomplete, but can instead be extended up to arbitrarily large $v$. In fact, null geodesics quickly converge towards an attractor at finite $r$ that we will investigate in more depth below.

As can be seen from the inset in Fig.~\ref{fig:nullgeonearrzeroreverse}, geodesics can start at small $r$ at finite $v$. Thus, it is not possible to set up an initial-value-problem in the spacetime, because there are always null geodesics to the future of any spatial hypersurface, which do not intersect the hypersurface when extended backward in time. This suggests that the spacetime should be extended to $r<0$. Numerically, null geodesics can be tracked into this region, but we do not explore this further.

\subsection{Energy conditions}
\label{subsec:QuantumEnergyConditions}
Given the radically altered behavior of null geodesics near $r=0$ that we observe numerically, we expect that energy conditions, which are prerequisites to prove geodesic incompleteness, may be violated in our spacetime. In fact, all pointlike energy conditions have counter-examples; e.g., in the form of classical scalar fields with non-minimal scalar curvature coupling  \cite{Bekenstein:1975ww,Deser:1983rq,Barcelo:2000zf,Barcelo:2002bv}  and in semiclassical gravity, see, e.g.,  \cite{Epstein:1965zza,Zeldovic:1971dx,Parker:1973qd,Roman:1986tp,Visser:1996iw,Visser:1996iv,Flanagan:1996gw,Visser:1997sd,Barcelo:2002bv, Kontou:2020bta}. We expect them to break down in any quantum gravity  inspired spacetime, see \cite{Kuipers:2019qby}  for an example in quantum gravity using effective-field-theory methods and \cite{Long:2020oma, Hossain:2005km} in Loop Quantum Gravity.
The violation of the pointwise energy conditions could then lead to geodesically complete spacetimes beyond GR. We focus on the NEC, because it is the weakest of the pointlike energy conditions, i.e., if it is violated, all other pointlike energy conditions are violated as well \cite{Visser:1995cc}. We leave aside the question whether averaged energy conditions, reviewed in \cite{Roman:2004xm,Curiel:2014zba,Kontou:2020bta}, hold or not.\\

The upgraded metric is actually a generalized Vaidya spacetime, because the upgrade results in a $v$- and $r$-dependence of the mass. Therefore, the effective energy-momentum tensor associated to the upgraded metric through the Einstein equations is that of a generalized Vaidya spacetime. 
This spacetime, understood as a solution of the Einstein equations, describes the gravitational collapse of shells of non-null fluid, additionally to shells of null dust present in the classical case, flowing radially towards the center in otherwise spherically symmetric, asymptotically flat, vacuum exterior spacetime.
Because in this paper this spacetime arises as an upgrade of the Vaidya spacetime, the additional component in the effective energy-momentum tensor is not due to non-null fluid in the spacetime, but rather encodes new-physics effects at an effective level. This is particularly useful when examining energy conditions.
The energy momentum tensor reads
\be
T^{\rm\, eff}_{\mu \nu} = \frac{\dot{m}[v,r]}{4 \pi r^2}
\, n_{\mu} n_{\nu} + \left(\frac{m'[v,r]}{4 \pi r^2} -\frac{m''[v,r]}{8\pi r}\right) \left(n_{\mu} l_{\nu} + n_{\nu} l_{\mu}\right) -\frac{m''[v,r]}{8 \pi r}\, g_{\mu \nu}
\label{eq:GeneralisedEnergyMomentumTensor}
\ee
with null vectors $n_{\mu} = \delta^0_{\mu},\, l_{\mu} = \frac{1}{2} \left(1 - \frac{2 G_0 m[v,r]}{r}\right) \delta^0_{\mu} - \delta^1_{\mu}$ satisfying $n_{\mu} l^{\mu} = -1$ \cite{Mkenyeleye:2014dwa}. Here, the dots correspond to differentiation with respect to $v$, while the primes refer to differentiation with respect to $r$.

\begin{figure}[h!]
\captionsetup[subfigure]{justification=Centering}
\begin{subfigure}[t]{0.45\linewidth}
\includegraphics[width=0.8\linewidth]{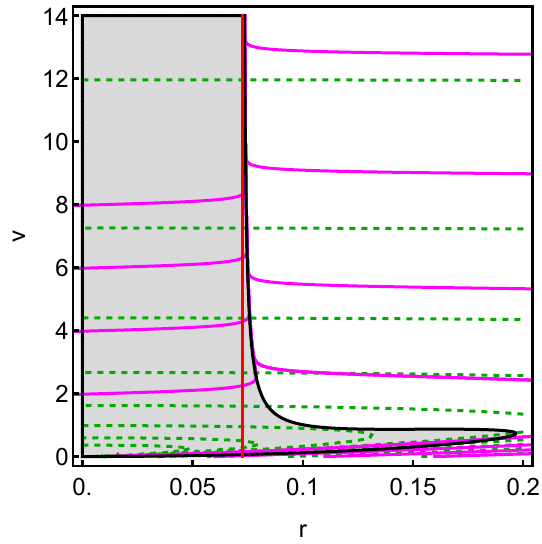}
\end{subfigure}\hspace{\fill}
\begin{subfigure}[t]{0.45\linewidth}
\includegraphics[width=\linewidth]{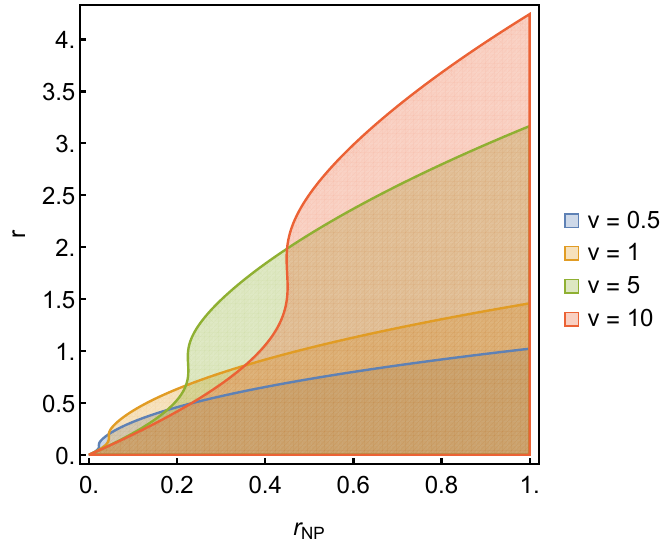}
\end{subfigure}

\caption{\label{fig:region_plots_NEC} Region plots of the violation of the NEC for arbitrary $k^0$, $G_0 \mu = \frac{1}{10}$, i.e., colored regions for which $\varepsilon < 0$. Left panel: region plot as a function of $v$ and $r$ for $r_{\rm NP} = \frac{39}{1000}$ together with classical (dashed green) and  upgraded  (magenta) null geodesics. Right panel: region plot as a function of $r_{\rm NP}$ and $r$ for $v = \frac{1}{2}$ (blue), $v = 1$ (orange), $v = 5$ (green) and $v = 10$ (red).}
\end{figure}

The null energy condition requires
\be
\varepsilon \geq 0
\ee
for the mass-energy density
\be
\varepsilon = T_{\mu \nu} k^{\mu} k^{\nu} = T_{00} (k^{0})^2 + 2\, T_{01} k^{0} k^{1} 
\label{eq:ModifiedNEC}
\ee
involving the energy momentum tensor $T_{\mu \nu}$ and a future-pointing null vector field $k^{\mu} = (k^0, k^1, k^2, k^3)$ which, by definition, satisfies
\be
k_{\mu} k^{\mu} = 0 \Leftrightarrow k^1 = \frac{1}{2} \left(1 - \frac{2 G_0 m[v, r]}{r}\right) k^0.
\label{eq:k1ask0}
\ee
We insert the upgraded mass function from Eq.~\eqref{eq:upgradedmassfunction}, focusing on the choice $n=1$ in the following. Substituting $k^1$ by its expression in terms of $k^0$ derived in Eq.~\eqref{eq:k1ask0} in the mass-energy density \eqref{eq:ModifiedNEC}, we obtain an expression in which $k^0$ only appears within a positive prefactor. 
Since we are interested in regions of spacetime and parameter space in which the NEC is violated, i.e., $\varepsilon < 0$, we can drop this positive prefactor and obtain the following lower bound 
\bea 
0 > &{}& r^7 + 8 \sqrt{3}\, G_0 r^4 r^2_{\rm NP} \mu v - 36 \sqrt{3}\, G_0 r^3 r^2_{\rm NP} v^2 \mu + 48\, G_0^2 r r^4_{\rm NP} v^2 \mu^2 \nonumber \\
&{}& + 72 \sqrt{3}\, G_0^2 r^2 r^2_{\rm NP} v^3 \mu^2 - 432\, G_0^2 r^4_{\rm NP} v^3 \mu^2.
\label{eq:ExplicitModifiedNEC}
\eea
This results in an upper bound on $r$, namely $r\lesssim r_{\rm NEC}[v, \mu, r_{\rm NP}]$.
Regions of negative mass-energy density, where the NEC is violated, are shown as colored shaded regions in Fig.~\ref{fig:region_plots_NEC} for an accretion rate of $\mu = \frac{1}{10 G_0} > \mu_{\rm c}$, where $\mu_{\rm c}$ is the critical accretion rate delineating the two classical regimes discussed in Sec.~\ref{sec:Classical_phase_diagram}. We observe that the attractor for null geodesics, that we already found in Sec.~\ref{subsec:quantnullgeodesicsnearcentre}, coincides with the boundary of the spacetime region in which the NEC is violated. The physics of the attractor may thus be understood as follows: for values $r > r_{\rm NEC}$, gravity acts as an attractive force and thus focuses geodesics. For values $r< r_{\rm NEC}$, the modification of the mass function implements a repulsive gravitational force, expelling null geodesics from this region and resulting in a violation of the NEC. As a consequence, the boundary of the region of violations of the NEC acts as an attractor for null geodesics.

For this rather large value of $\mu$, the last two terms in Eq.~\eqref{eq:ExplicitModifiedNEC} dominate and lead to a constant upper bound $r_{\rm NEC} =  3^{\frac{1}{4}} \sqrt{2} r_{\rm NP}$ for $r$ at a given value of $r_{\rm NP}$, represented by the red straight line in the left panel of Fig.~\ref{fig:region_plots_NEC}. At small values of $v$, the other terms in Eq.~\eqref{eq:ExplicitModifiedNEC} start playing a role and are responsible for the tail to larger values of $r$ visible in the left panel of Fig.~\ref{fig:region_plots_NEC}. The behavior of $\varepsilon$ as a function of $r$ and $r_{\rm NP}$ in the right panel of Fig.~\ref{fig:region_plots_NEC} interpolates between a linear regime at very small values of $r_{\rm NP}$ given by the contributions of the last two terms in Eq.~\eqref{eq:ExplicitModifiedNEC}, and a quadratic polynomial at larger values of $r_{\rm NP}$ due to the contributions of additional terms.

\begin{figure}[h!]
\captionsetup[subfigure]{justification=Centering}
\begin{subfigure}[t]{0.45\linewidth}
\includegraphics[width=0.8\linewidth]{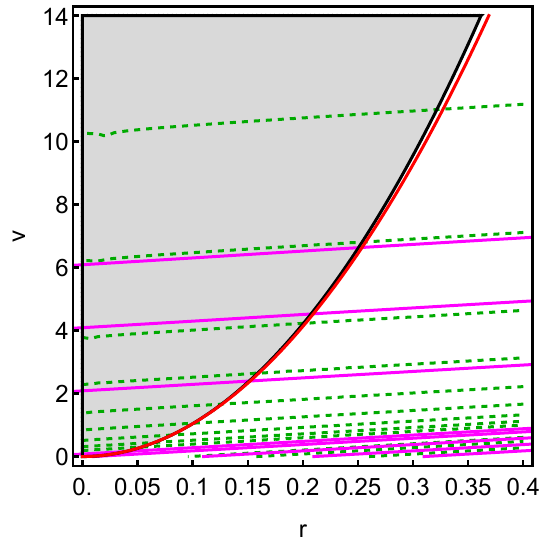}
\end{subfigure}\hspace{\fill}
\begin{subfigure}[t]{0.45\linewidth}
\includegraphics[width=\linewidth]{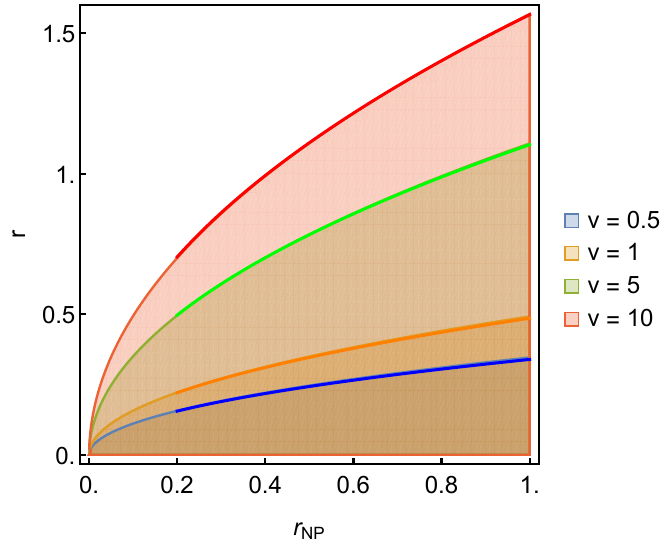}
\end{subfigure}
\caption{\label{fig:region_plots_NEC_Small_Mu} Region plots of the violation of the NEC for arbitrary $k^0$, $G_0 \mu = \frac{1}{1000}$, i.e., colored regions for which $\varepsilon < 0$. Left panel: region plot as a function of $v$ and $r$ for $r_{\rm NP} = \frac{39}{1000}$  together with classical (dashed green) and upgraded (magenta) null geodesics. Right panel: region plot as a function of $r_{\rm NP}$ and $r$ for $v = \frac{1}{2}$ (blue), $v = 1$ (orange), $v = 5$ (green) and $v = 10$ (red).}
\end{figure}

We find a qualitative difference in the shape of the regions where the NEC is violated for small accretion rates, e.g., $G_0 \mu = \frac{1}{1000} \ll G_0 \mu_{\rm c}$, in Fig.~\ref{fig:region_plots_NEC_Small_Mu}. This stems from the fact that, when $\mu$ is very small, the first three terms in Eq.~\eqref{eq:ExplicitModifiedNEC} dominate.
Indeed, considering only the first three terms in Eq.~\eqref{eq:ExplicitModifiedNEC} and solving for $r[v]$, one obtains a very good approximation of the outer boundary of the gray shaded region in the left panel of Fig.~\ref{fig:region_plots_NEC_Small_Mu}, represented by a red curve. In a similar way, we can find $r_{\rm NEC}[r_{\rm NP}]$ for every chosen value of $v$ in the right panel of Fig.~\ref{fig:region_plots_NEC_Small_Mu}, and the resulting curves delineate the outer boundary of the shaded regions where the NEC is violated. As displayed in the left panel of Fig.~\ref{fig:region_plots_NEC_Small_Mu}, the motion of geodesics is not correlated with the region in which the NEC is violated, because, in the classical horizonless case, i.e., $G_0 \mu = \frac{1}{1000}$, null geodesics already escape from $r = 0$ and this behavior does not change when the spacetime is upgraded.\\

\subsection{Spacetime diagrams}
\label{subsec:spacetimediagrams}
To construct spacetime diagrams of the upgraded spacetime, we have to find the special surfaces of the spacetime, namely its horizons as well as the distinguished photon surface, in the absence of a photon sphere.

\subsubsection{Inner, apparent and event horizons}
As shown in App.~\ref{app:DefEqApparentHorizon}, the location of the apparent horizon in the upgraded spacetime is found by solving the equation $g^{rr}_{\rm EF} = 0$. This equation generically admits two real solutions. The first, given for the mass function \eqref{eq:upgradedmassfunction} with $n = 1$ by 
\bea
r_{\rm AH} &=& \frac{1}{3} \Bigg[ 2 G_0 \mu v + \frac{2\cdot 2^{\frac{2}{3}}\left(G_0 \mu v\right)^{\frac{5}{3}}}{\left(4 G_0^2 \mu^2 v^2 +3 \sqrt{3}\,r_{\rm NP}\left(- 9\, r_{\rm NP} +\sqrt{81 r_{\rm NP}^2 - 8 \sqrt{3}\, G_0^2 \mu^2 v^2}\right) \right)^{\frac{1}{3}}}\nonumber\\
&{}&+ \left(8 G_0^3 \mu^3 v^3 + 6 \sqrt{3}\,r_{\rm NP} G_0 \mu v \left(- 9 r_{\rm NP}+ \sqrt{81 r_{\rm NP}^2 - 8 \sqrt{3}\, G_0^2 \mu^2 v^2}\right)\right)^{\frac{1}{3}}
\Bigg],
   \label{eq:GeneralisedApparentHorizon}
\eea
is the apparent horizon. As discussed in Sec.~\ref{subsubsec:simplicity}, the upgraded collapsing object is more compact than its classical counterpart, which one can infer from the additional new-physics terms proportional to $r_{\rm NP}$ present in the denominator of Eq.~\eqref{eq:GeneralisedApparentHorizon}.\\
The second solution
\bea
r_{\rm IN} &=& \frac{1}{3} \Bigg[2 G_0 \mu v  + \frac{2 \cdot 2^{-\frac{1}{3}} (-1+i\sqrt{3}) (G_0 \mu v)^{\frac{5}{3}}}{\left(4 G_0^2 \mu^2 v^2 + 3 \sqrt{3} r_{\rm NP} \left(-9 r_{\rm NP} + \sqrt{81
   r_{\rm NP}^2-8 \sqrt{3} G_0^2 \mu ^2 v^2}\right)\right)^{\frac{1}{3	}}}\nonumber\\
&{}&+\left(-1-i \sqrt{3}\right) \left(8 G_0^3 \mu^3 v^3 + 6 \sqrt{3} r_{\rm NP} G_0 \mu v \left(-9 r_{\rm NP} + \sqrt{81 r_{\rm NP}^2-8 \sqrt{3} G_0^2 \mu ^2 v^2}\right)\right)^{\frac{1}{3}}\!\!\!\Bigg]
\label{eq:GeneralisedInnerHorizon}
\eea
is a new, inner apparent horizon. This expression
is real up to the critical point $r_{\rm NP} = r_{\rm NP, crit}$, after which it becomes complex. The inner horizon appears, because the modification function $f_{\rm NP}[I_1 r^4_{\rm NP}]$ entering Eq.~\eqref{eq:EquationApparentInnerHorizons} that determines the location of horizons is very small at small $r$, such that the equation $g_{\rm EF}^{rr} = 0$ accommodates another solution. This behavior is well-known for stationary regular black-hole spacetimes, see, e.g., the classic examples \cite{Dymnikova:1992ux, Hayward:2005gi} and can be argued for on general grounds \cite{Carballo-Rubio:2019fnb,Carballo-Rubio:2022nuj,Carballo-Rubio:2023mvr}.

Solutions to algebraic equations become complex in pairs and thus at $r_{\rm NP, crit}$, the outer and inner apparent horizons merge and subsequently become complex, leaving behind a horizonless spacetime. These results mirror those obtained for the event horizon in the stationary limit with a constant mass
\cite{Eichhorn:2022oma}.

\begin{figure}[!t]
\captionsetup[subfigure]{justification=Centering}
\begin{subfigure}[t]{0.45\linewidth}
\includegraphics[width=\linewidth]{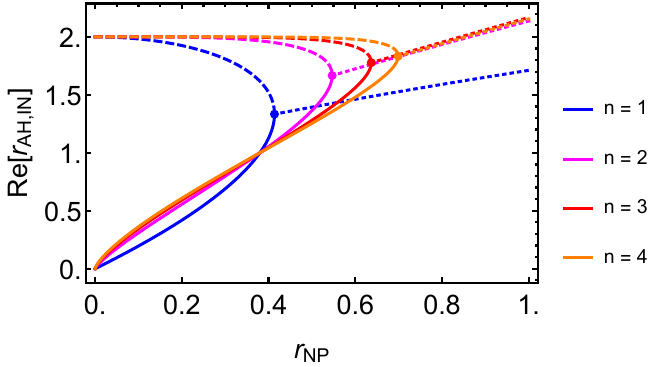}
\caption{\label{subfig:Re_rAH_n_v_small}
${\rm Re}[r_{\rm AH, IN}]$ for $\mu = \frac{1}{10}$, $v = 10$.}
\end{subfigure}\hspace{\fill}
\begin{subfigure}[t]{0.45\linewidth}
\includegraphics[width=\linewidth]{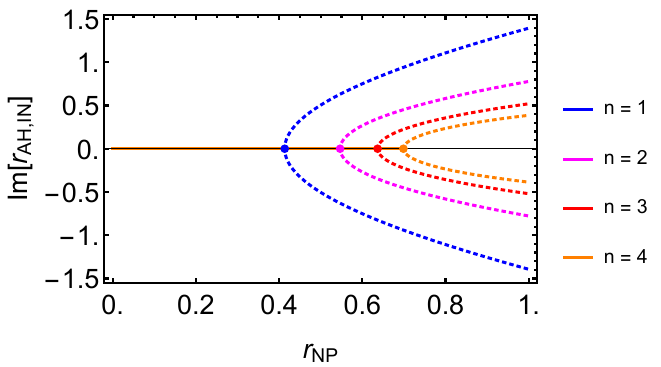}
\caption{\label{subfig:Im_rAH_n_v_small}
${\rm Im}[r_{\rm AH, IN}]$ for $\mu = \frac{1}{10}$, $v = 10$.}
\end{subfigure}\vspace{\fill}
\begin{subfigure}[t]{0.45\linewidth}
\includegraphics[width=\linewidth]{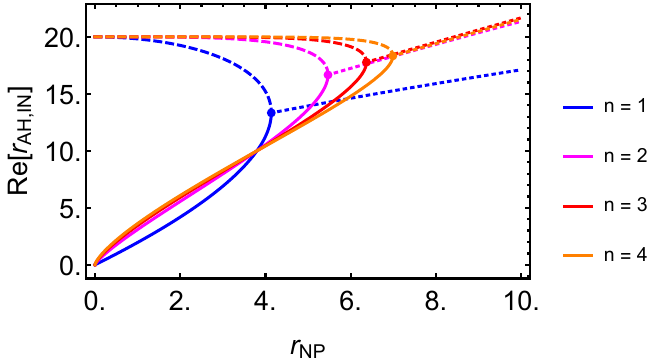}
\caption{\label{subfig:Re_rAH_n_v_big}
${\rm Re}[r_{\rm AH, IN}]$ for $\mu = \frac{1}{10}$, $v = 100$.}
\end{subfigure}\hspace{\fill}
\begin{subfigure}[t]{0.45\linewidth}
\includegraphics[width=\linewidth]{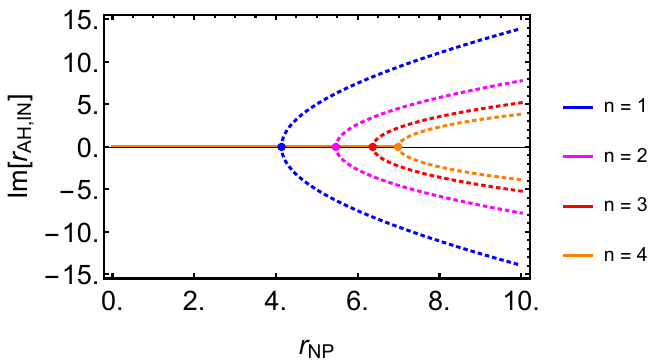}
\caption{\label{subfig:Im_rAH_n_v_big}
${\rm Im}[r_{\rm AH, IN}]$ for $\mu = \frac{1}{10}$, $v = 100$.}
\end{subfigure}

\caption{\label{fig:rAH_n} Left column: real part of $r_{\rm AH}$ (dashed line) and $r_{\rm IN}$ (solid line) as a function of $r_{\rm NP}$ for $n = 1, 2, 3, 4$. Right column: imaginary part of $r_{\rm AH}$ and $r_{\rm IN}$ as a function of $r_{\rm NP}$ for $n = 1, 2, 3, 4$. The parameters values for Figs.~\ref{subfig:Re_rAH_n_v_small}, \ref{subfig:Im_rAH_n_v_small} are $0\leq r_{\rm NP}\leq 1$ , $G_0 \mu = \frac{1}{10}$, $G_0 = 1$ and $v = 10$, while they are $0\leq r_{\rm NP}\leq 10$ , $G_0 \mu = \frac{1}{10}$, $G_0 = 1$ and $v = 100$ for Figs.~\ref{subfig:Re_rAH_n_v_big}, \ref{subfig:Im_rAH_n_v_big}. The colored points indicate the locations of the critical points $r_{\rm NP, crit, n}$.}
\end{figure}

Fig.~\ref{fig:rAH_n} shows both the real and imaginary parts of the two solutions to $g_{\rm EF}^{rr} = 0$ as a function of $r_{\rm NP}$, for the different powers $n = 1, 2, 3, 4$ in the mass function Eq.~\eqref{eq:upgradedmassfunction}.
For each power $n$, there is, as expected, one solution which decreases with $r_{\rm NP}$ until the critical point $r_{\rm NP, crit, n}$ and corresponds to the apparent horizon; and another solution which increases with $r_{\rm NP}$, corresponding to a (new) inner horizon. They merge at the critical point where they become complex. Beyond the critical point, the two solutions acquire an imaginary part, rendering the spacetime horizonless.

As opposed to the stationary case, the critical value of the new-physics scale $r_{\rm NP, crit, n}$ is not a constant but increases linearly with $v$,
\be
r_{\rm NP, crit, n} \equiv r_{\rm NP, crit, n}[v] = c_n\, \mu v,\quad c_n = {\rm const.}\: \forall n. 
\ee
Thus, as $v$ is taken to very small values, $r_{\rm NP, crit, n}$ approaches zero. Therefore,
the formation of the apparent horizon is delayed by the new-physics effects. This is because for early enough times i.e., small enough $v$, $r_{\rm NP}$ (which is a fixed constant) is always larger than $r_{\rm NP, crit}$ (which grows linearly in $v$). Accordingly, the apparent horizon cannot exist for these very early times and its formation can only proceed once $r_{\rm NP} < r_{\rm NP, crit}[v]$. 

This effect is of particular interest in situations where the classical spacetime exhibits a naked singularity. New-physics effects resolve the singularity and limit the maximum value of curvature invariants. The spacetime region which is thus affected is not hidden behind an apparent horizon at early times, given that  horizon-formation is delayed by the same new-physics effects. An asymptotic observer may therefore (in principle) access this spacetime region through observations. This is an interesting distinction to the stationary case, where $r_{\rm NP, crit}$ is a constant and there are thus always choices of $r_{\rm NP} < r_{\rm NP, crit}$ which result in large modifications of the spacetime being hidden behind its event horizon.\\

\begin{figure}[h!]
\centering
\includegraphics[width=0.75\linewidth]{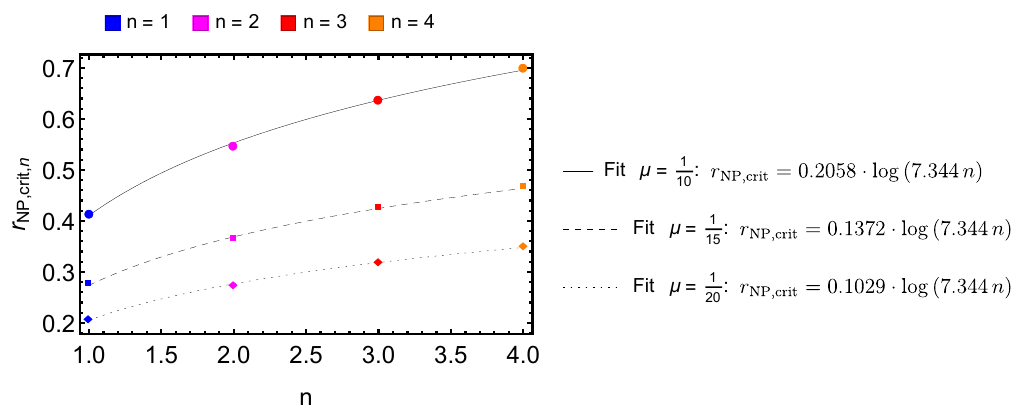}
\caption{\label{fig:rNP_crit_n}
We show the critical points $r_{\rm NP, crit, n}$ as a function of $n$ for $v = 10$ with either $G_0 \mu = \frac{1}{10}$ (continuous line), $G_0 \mu = \frac{1}{15}$ (dashed line) or $G_0 \mu = \frac{1}{20}$ (dotted line). Logarithmic fits haven been displayed to guide the eyes of the reader.}
\end{figure}

The dependence of $r_{\rm NP, \, crit, n}$ on $n$ is non-linear, cf.~Fig.~\ref{fig:rNP_crit_n}. The reason is that increasing $n$ in the upgraded mass function in Eq.~\eqref{eq:upgradedmassfunction} leads to a faster approach of $f_{\rm NP}[I_1 r^4_{\rm NP}] \rightarrow 1$. Thus, the relative modification of the spacetime is smaller at the classical location of the apparent horizon, the larger $n$ is, and the location of the inner horizon is closer to $r=0$ for larger $n$. 
Therefore, the merging of the two horizons is delayed to larger $r_{\rm NP}$ for larger $n$.

As in the classical case, in addition to the apparent horizon, there is also an event horizon.
To determine its location $r_{\rm EH}[v]$, we use that for $v \geq \bar{v}$ the spacetime is isometric to a modified Schwarzschild black hole with  constant mass 
\be
m \equiv \frac{\mu \bar{v}}{r_{\rm EH}[\bar{v}]\, +\, r^{2n}_{\rm NP} \left(48\, G^2_0\, \mu^2\, \bar{v}^2\, r_{\rm EH}[\bar{v}]^{\left(\frac{2}{n} - 6\right)}\right)^{\frac{n}{2}}},
\label{eq:MassModifiedSchwarzschild}
\ee
 and thus an event horizon located at a constant radius $r_{\rm EH}[\bar{v}]$, that means $ \dot{r}_{\rm EH}[v] = 0$.
In practice, we first determine the initial condition by solving $1 - \frac{2 G_0 m}{r_{\rm EH}[\bar{v}]} = 0$ for  $r_{\rm EH}[\bar{v}]$ with constant mass $m$ given in Eq.~\eqref{eq:MassModifiedSchwarzschild}. We then numerically solve the geodesic equation for $r_{\rm EH}[v]$ backward in  time, i.e. for $0 < v < \bar{v}$, together with the initial conditions $r_{\rm EH}[\bar{v}]$ and $\dot{r}_{\rm EH}[v] = 0$ at $v > \bar{v}$. 

\subsubsection{Loss of the photon sphere and distinguished photon surface}
\label{subsec:Quantum_Null_Geodesic_Motion}
As briefly mentioned in Sec.~\ref{sec:Photon sphere}, the photon sphere that exists in the classical spacetime for subcritical accretion rates $\mu < \mu_{\rm c}$ is lost in the upgraded spacetimes. This stems from the loss of the conformal Killing vector when the dynamical mass function acquires a non-linear dependence on $v$ and/or an additional dependence on $r$, see \cite{nielsen2014revisiting}. Hence, it is a generic feature of generalised Vaidya spacetimes with dynamical mass functions depending both on $v$ and $r$, and thus a feature of the upgraded spacetimes with mass functions Eq.~\eqref{eq:upgradedmassfunction}.

As a result, we instead focus on photon surfaces which exist for all values of $\mu$ and for both the classical and the upgraded spacetimes. Among all photon surfaces, we choose to dinstinguish the unique photon surface that matches onto the photon sphere of the modified Schwarzschild black hole at $v > \bar{v}$ and trace it back at $0 \leq v < \bar{v}$. This choice allows us to readily compare the distinguished photon surface in the upgraded spacetimes with that in the classical VKP spacetime.\\
Following \cite{Mishra:2019trb}, we
write the equation for the location of any photon surface for a general mass function $m[v,r]$ as\footnote{As previously, we define $\dot{m} = \frac{\partial m[v,r]}{\partial v}$ and $m' = \frac{\partial m[v,r]}{\partial r}$.}
\bea
0&=&\ddot{r}_{\rm p}[v] + \frac{1}{r_{\rm p}[v]} \left(3 \dot{r}_{\rm p}[v] + G_0 m' (3 \dot{r}_{\rm p}[v] - 1) - 2 \left(\dot{r}_{\rm p}[v]\right)^2 + G_0 \dot{m} - 1\right)\nonumber\\
&{}& + \frac{1}{r_{\rm p}^2 [v]} \left(- 9 G_0 m \dot{r}_{\rm p}[v] + 2 G_0^2 m m' + 5 G_0 m\right) - \frac{6 G_0^2 m^2}{r_{\rm p}^3[v]}.
\label{eq:QuantEquationPhotonSphere}
\eea
We do not find an analytical solution to this equation and hence solve it numerically. To obtain the distinguished photon surface, we supplement the differential equation with the appropriate initial conditions at $v > \bar{v}$. As the collapsing compact object settles down to a modified Schwarzschild (thus static) black hole starting from $v = \bar{v}$, with a photon sphere located at a fixed radius, the  initial conditions are given by the largest real solution to the following equation (see Eq.~(\textcolor{blue}{2.21}) of \cite{Eichhorn:2022oma})
\be
1 - \frac{3 G_0 \mu \bar{v}}{r_{\rm p}[\bar{v}]} f_{\rm NP}[I_1 \cdot r_{\rm NP}^4]\bigg\vert_{r = r_{\rm p}[\bar{v}]} + G_0 \mu \bar{v} \frac{\partial f_{\rm NP}[I_1 \cdot r_{\rm NP}^4]}{\partial r}\bigg\vert_{r = r_{\rm p}[\bar{v}]} = 0,
\label{eq:QuantBCPhotonSphere}
\ee
and the requirement that $\dot{r}_{\rm p}[v] = 0$ for $v \geq \bar{v}$. While  finding the location of the distinguished photon surface $r_{\rm p}[v]$ for $0 < v < \bar{v}$ now seems straightforward, there is a remaining caveat: the upgraded mass function in Eq.~\eqref{eq:upgradedmassfunction} has the same sharp transition at $v = \bar{v}$ as the (classical) VKP mass function described in Eq.~\eqref{eq:VKPmass}, which can cause difficulties in the numerical solution. We instead implement a smooth approximation to the full mass function by using the same approximation  for $m[v]$ as in  the classical case Eq.~\eqref{eq:SmoothApproxVKPmass}.
The resulting upgraded distinguished photon surface is displayed in magenta in Figs.~\ref{fig:PhaseDiagQuantNullGeon1} and \ref{fig:PhaseDiagQuantNullGeon1SmallMu} for $G_0 \mu = \frac{1}{10}$ and $G_0 \mu = \frac{1}{20}$, respectively. For both cases, the accretion rate is low enough that the distinguished photon surface initially starts at a null surface of the type $v = {\rm const.}$ together with the matter shells,  moves inwards before reaching its final radial location given by a real solution of Eq.~\eqref{eq:QuantBCPhotonSphere}. The difference lies in the initial value of $v$ at which the distinguished photon surface forms, which is smaller for larger accretion rates as the collapsing body is more compact at earlier times. Comparing with the classical case in Fig.~\ref{fig:PhaseDiagClassNullGeo}, the behavior of the upgraded distinguished photon surface is similar. This is likely due to the fact that new-physics modifications impact the region well inside the apparent horizon, while leaving the distinguished photon surface unaffected, c.f.~Fig.~\ref{fig:ComparisonClassQuantNullGeon1}, apart from its final location at $v = \bar{v}$. Furthermore, the distinguished photon surface for the horizonless spacetime only exists for a limited range of $r^1_{\rm NP, crit} < r_{\rm NP} < r^2_{\rm NP, crit}$, as was shown for the stationary case in \cite{Eichhorn:2022oma}. If $r_{\rm NP}$ is too large, then the distinguished photon surface and its associated inner counterpart merge and the solutions to the differential equation Eq.~\eqref{eq:QuantEquationPhotonSphere} all become complex.

\subsubsection{Spacetime diagrams}
\begin{figure}[h!]
\centering
\subfloat{\includegraphics[width=0.45\linewidth]{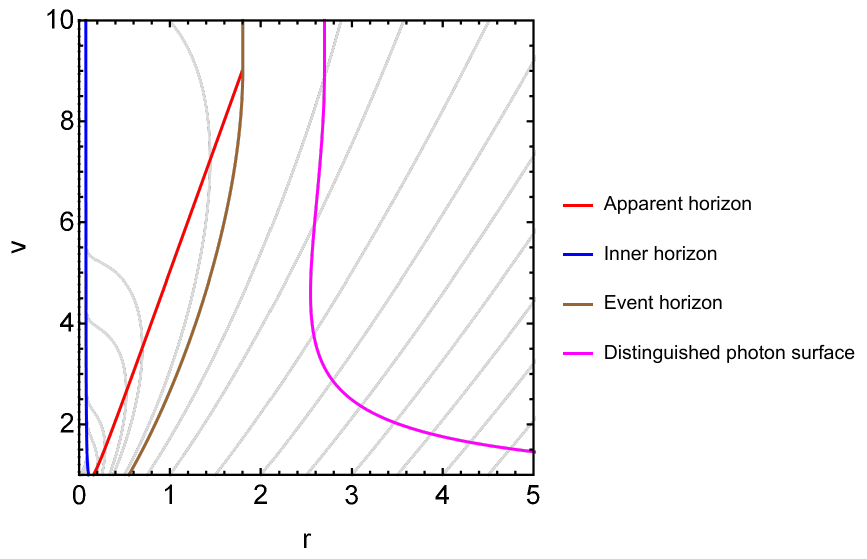}}
\subfloat{\includegraphics[width=0.44\linewidth]{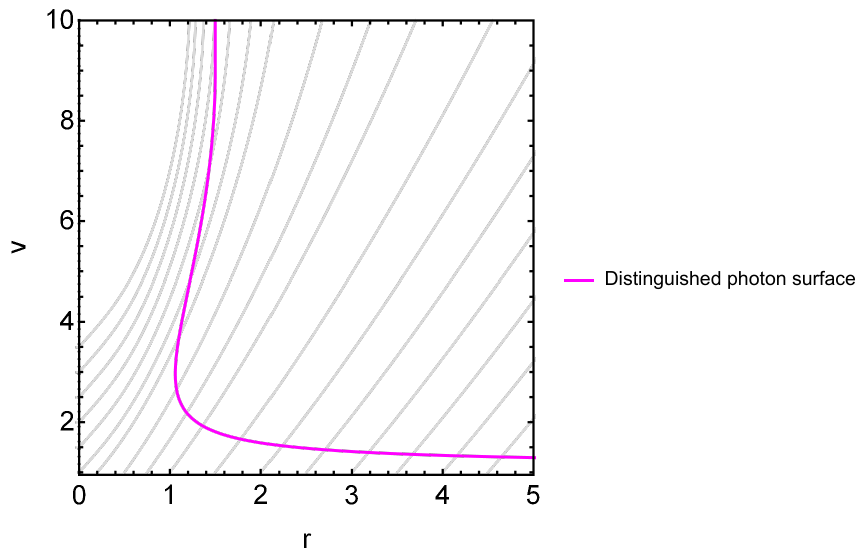}} 
\caption{Spacetime diagrams of
$(r[v], v)$ of null geodesics with $n = 1$ and $G_0 \mu = \frac{1}{10}$. Left panel: case with a horizon, i.e. $r_{\rm NP} < \min\limits_{v}{(r_{\rm NP, crit, 1} [v])}$. Right panel: horizonless case, i.e., $r_{\rm NP} > \max\limits_{v}{(r_{\rm NP, crit, 1} [v])}$.
 \label{fig:PhaseDiagQuantNullGeon1}}
\end{figure}

\begin{figure}[h!]
\centering
\subfloat{\includegraphics[width=0.45\linewidth]{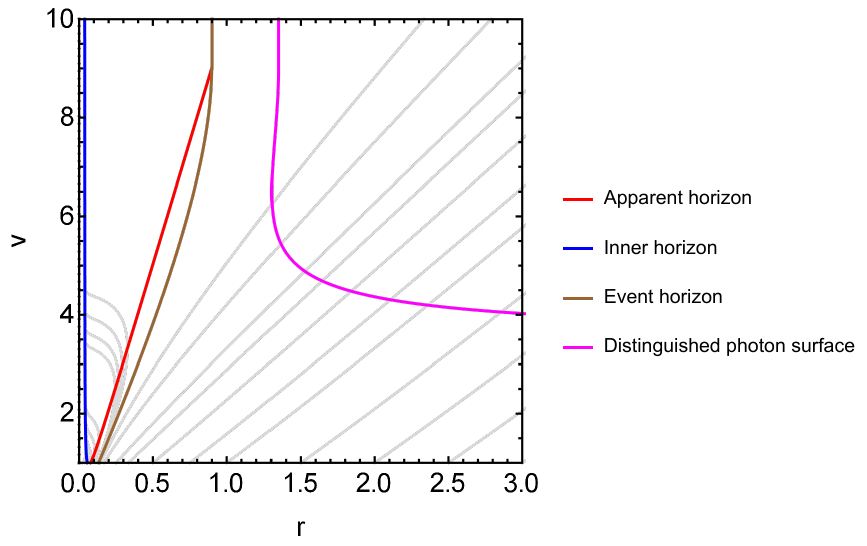}}
\subfloat{\includegraphics[width=0.45\linewidth]{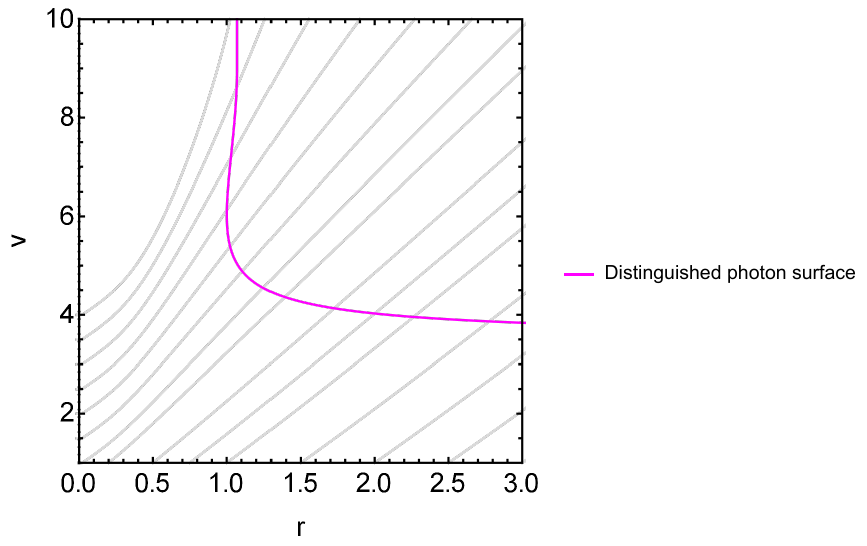}} 
\caption{Spacetime diagrams of $(r[v], v)$ for $n = 1$ and $G_0 \mu = \frac{1}{20}$. Left panel: case with an horizon, i.e. $r_{\rm NP} < \min\limits_{v}{(r_{\rm NP, crit, 1} [v])}$. Right panel: horizonless case, i.e. $r_{\rm NP} > \max\limits_{v}{(r_{\rm NP, crit, 1} [v])}$.
\label{fig:PhaseDiagQuantNullGeon1SmallMu}}
\end{figure}

The spacetime diagrams that summarize our analysis are shown in Figs.~\ref{fig:PhaseDiagQuantNullGeon1} (for $G_0 \mu = \frac{1}{10}$) and \ref{fig:PhaseDiagQuantNullGeon1SmallMu} (for $G_0 \mu = \frac{1}{20}$). 

For each value of the accretion rate, we study two distinct regimes that correspond to different choices of $r_{\rm NP}$: the first one corresponds to a spacetime with an event horizon (left panel), while the second one is horizonless (right panel). Since the presence of an event horizon depends on the ratio $\frac{r_{\rm NP}}{r_{\rm NP, crit, 1}[v]}$, the above statement can be reformulated as follows: the upgraded spacetime has an event horizon as long as $r_{\rm NP} < \min\limits_{v}{\left(r_{\rm NP, crit, 1}[v]\right)},\, 1 \leq v \leq 10$, and is horizonless if $r_{\rm NP} > \max\limits_{v}{\left(r_{\rm NP, crit, 1}[v]\right)},\, 1 \leq v \leq 10$.

\begin{figure}[!t]
\centering
\subfloat{\includegraphics[width=0.43\linewidth]{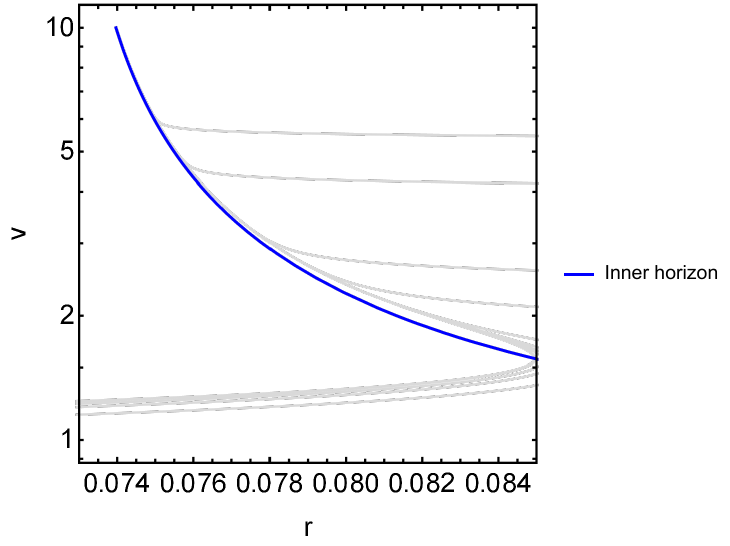}} 
\subfloat{\includegraphics[width=0.46\linewidth]{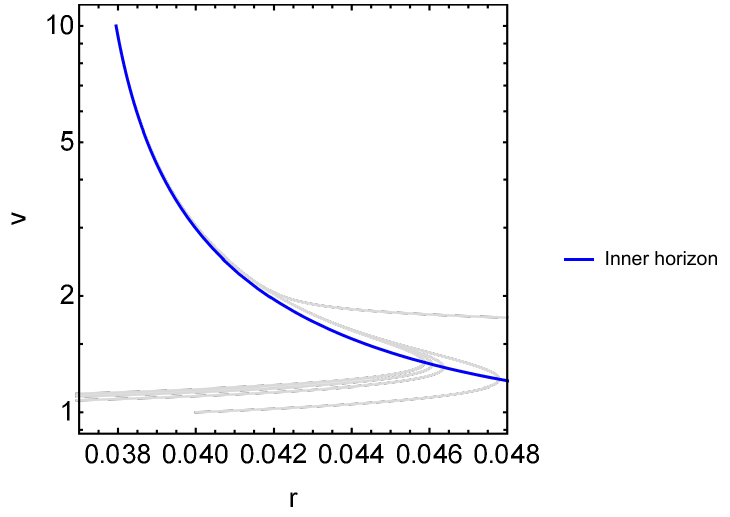}}  
\caption{Zoom-in on the spacetime diagram of $(r[v], v)$ of null geodesics to small $r$ with $n = 1$ for $G_0 \mu = \frac{1}{10}$ (left panel) and $G_0 \mu = \frac{1}{20}$ (right panel). For both panels, the scale is logarithmic in $v$ and there is an horizon, i.e., $r_{\rm NP} < \min\limits_{v}{(r_{\rm NP, crit, 1} [v])}$.
\label{fig:ZoomsPhaseDiagQuantNullGeon1}}
\end{figure}
In the spacetime diagrams, it is again apparent that there is an attractor for null geodesics inside the apparent horizon. Numerically, this attractor is initially close to, but not in agreement with, the inner horizon, before converging to the inner horizon at larger $v$, cf.~Fig.~\ref{fig:ZoomsPhaseDiagQuantNullGeon1}. Note that the attractor converges faster to the inner horizon for smaller values of $G_0 \mu$. Hence, the inner horizon is acting as an inner trapping horizon.

This, in turn, likely causes a problem: the energy carried by photons trapped on the attractor accumulates, leading the spacetime curvature to rise\footnote{This of course assumes that the GR equations of motion continue to hold -- new-physics effects that limit the maximum value of curvature invariants in the spacetime and that are implemented through the regularity principle would be expected to alter this response of the spacetime metric to an increase in the energy-momentum tensor, at least beyond a critical value of the energy/momentum.} and triggering a potential instability \cite{DiFilippo:2022qkl}. Because we do not account for backreaction in our analysis, this effect is not visible in our spacetime diagrams. Future investigations of this point will be crucial to determine whether or not the upgraded Vaidya spacetime can stay regular also in the fully dynamical case. The outcome of such an analysis of course depends on the assumed dynamics; the intuition that an attractor for geodesics may result in the build-up towards a spacetime singularity may not hold in dynamics beyond GR.

However, the attractor also has a desired consequence, namely that it solves the predictivity problem connected to geodesics that emanate at $(r=0, v > 0)$. These geodesics are not past complete within the spacetime region with $r \in [0,\infty)$. Thus, they pose a problem with setting up an initial-value problem. However, because those geodesics get trapped by the attractor, it effectively shields the external spacetime from a breakdown of predictivity. These comments, of course, only apply to radial geodesics and we do not investigate more general null geodesics here.

\begin{figure}[h!]
\centering
\includegraphics[width=0.7\linewidth]{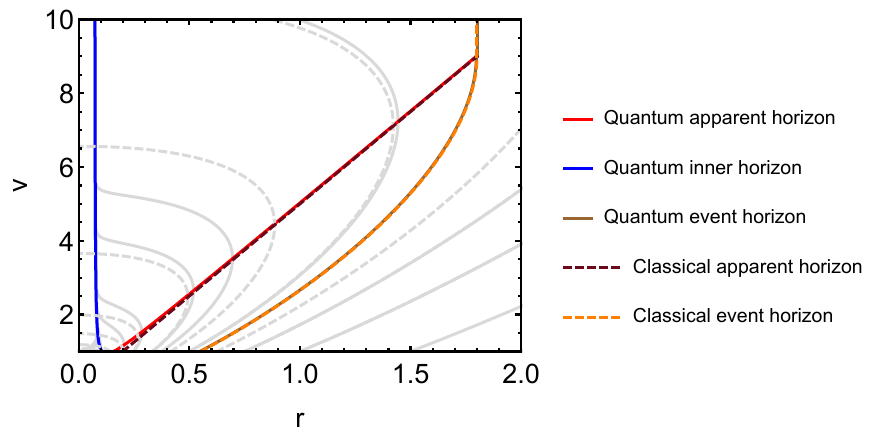}
\caption{Spacetime diagram of $(r[v], v)$ for radial null geodesics in both the classical VKP model (dashed lines) and the upgraded Vaidya model with $n = 1$ (continuous lines), in the presence of an horizon, i.e. $r_{\rm NP} < \min\limits_{v}{(r_{\rm NP, crit, 1} [v])}$.
\label{fig:ComparisonClassQuantNullGeon1}}
\end{figure}

A comparison of the behavior of upgraded and classical null geodesics near the center is shown in Fig.~\ref{fig:ComparisonClassQuantNullGeon1}. As expected, the largest deviations occur for relatively small $r$, whereas the apparent and event horizons of the classical and the upgraded spacetime already lie nearly on top of each other.

\section{Outlook:  regular gravitational collapse beyond the Vaidya model}
\label{sec:Outlook}
In this paper, we have extended the principled-parameterized approach to black-hole spacetimes beyond GR from the stationary case \cite{Eichhorn:2021iwq,Eichhorn:2021etc,Eichhorn:2022oma} to the time-dependent setting of gravitational collapse. We have focused on a simple model and now  sketch how the approach may also be used in less simple models, e.g., in the generalized Vaidya spacetime. In our paper, the framework of generalized Vaidya spacetimes has been used to describe the \emph{endpoint} of the upgrade procedure. Now we ask, whether it can also be used as the \emph{starting} point. To that end, we need to consider the curvature invariants, because a suitable measure of local curvature is the key ingredient in the principled-parameterized approach.

Among the full set of the 17 ZM invariants in App.~\ref{app:GeneralisedZMinvariants} for the generalized Vaidya spacetime, ten are non-zero. Most of those non-zero invariants cannot be written as powers of each other, unless it is explicitly indicated how to do so in App.~\ref{app:GeneralisedZMinvariants}. However, this does not mean that the remaining eight non-zero invariants are all algebraically independent of each other. 
Syzygies, which are polynomial relations between the
curvature invariants, may be found. We  find a nontrivial syzygy, namely
\be
\frac{I_{11}}{I_1} + \frac{I_5^2}{12} - \frac{I_6}{3} = 0
\label{eq:ExampleSyzygy}.
\ee 
This reduces the number of polynomially independent curvature invariants to be at most six.\\

We postulate that, if the modulus of any of the non-zero, independent curvature invariants exceeds a critical value, modifications to the spacetime become sizable. We thus construct a measure of the local curvature as follows: we take the absolute value of each classical curvature invariant to an appropriate power, so that it has the same dimensionality as the Kretschmann scalar $I_1$. We then form the average of these quantities, i.e., the root-mean-square (RMS) of the sum of appropriate powers of the classical curvature invariants. By using the RMS, we avoid a bias with respect to the sign of the curvature, assuming that new physics kicks in at large positive or negative curvatures. This amounts to considering a local curvature $K$ of the general form
\be
K = \frac{1}{N} \sqrt{\sum_{j = 1}^{N} | I_j |^{\alpha_j}},\quad \alpha_j \in \mathbb{Q},
\label{eq:FormulaLocalCurvatureScale}
\ee
with $I_j$ being the non-zero, polynomially independent, curvature invariants of the classical non-upgraded spacetime.\\

Using only the independent curvature invariants, the local curvature scale of the generalized Vaidya spacetime takes the form
\be
K = \frac{1}{6} \sqrt{\vert I_1 \vert + \vert I_5 \vert^2 + \vert I_6 \vert + \vert I_7 \vert^{\frac{2}{3}} + \vert I_8 \vert^{\frac{1}{2}} + \vert I_9 \vert^{\frac{2}{3}} + \vert I_{13} \vert^{\frac{2}{5}}},
\label{eq:GeneralisedLocalCurvatureScale}
\ee
where the implicit dependence of $K$ in $m[v,r]$ and its $r$-derivatives $m'[v,r]$, $m''[v,r]$ is hidden in the curvature invariants.

In practice, this expression is significantly more complicated than simply $I_1$, which we could use for the Vaidya spacetime. Thus, the principled-parameterized approach should be applicable to the generalized Vaidya spacetime, but we expect the numerical analysis of the spacetime to be more involved than in the present case.

\section{Conclusions}
\label{sec:Conclusion}
In this paper, we apply the principled-parameterized approach \cite{Eichhorn:2021etc,Eichhorn:2021iwq,Eichhorn:2022oma} to spherically symmetric gravitational collapse described by a Vaidya metric. In this framework, new-physics effects - for instance motivated by quantum-gravity theories such as asymptotically safe gravity - are parameterized by an upgraded mass function that depends on the curvature invariants of the classical spacetime and thereby the spacetime coordinates. 

Our main results are the following:
\begin{enumerate}
\item[a)] The modification functions $f_{\rm NP}$, that modify the time-dependent mass term and implement the four principles underlying the principled-para\-meterized approach for the time-dependent gravitational collapse, are the same modification functions that have been found in the stationary \cite{Eichhorn:2021etc,Eichhorn:2021iwq} as well as the static case 
\cite{Eichhorn:2022oma}. This suggests a \emph{universality} of the modification function in the sense that the same function can be used to render rather distinct black-hole spacetimes regular. The modification function automatically adapts itself to the properties of a given spacetime, because it depends on the spacetime's curvature invariants.\\ 
In the future, this calls for a study of other models of gravitational collapse within the principled-parameterized approach.\\
A universality of the modification function arises naturally in asymptotic-safety inspired spacetimes which are constructed by RG improvement. There, it is actually not the mass itself that is upgraded to a spacetime-dependent function, but rather the Newton constant which always appears as a multiplicative factor in front of the mass. RG improvement has been explored extensively, see the reviews \cite{Eichhorn:2022bgu,Platania:2023srt}. We argue that many scale identifications that have been used do not conform with a notion of local coarse-graining underlying the RG, e.g., a scale identification with the geodesic distance from the black hole's center is clearly non-local. Instead, the local value of curvature invariants\footnote{For spacetimes with few or none Killing vectors, several non-derivative curvature invariants are typically polynomially independent. In this case, a useful notion of local curvature consists in the RMS of all independent invariants.} constitutes a local interpretation of the RG scale. Because this scale interpretation renders all black-hole spacetimes that have been explored to date regular (including static, stationary as well as spherically symmetric, time-dependent), we conjecture that it is a preferred scale interpretation.
\item[b)] Implementing a mass function that satisfies regularity, i.e., the absence of curvature singularities, generates a spacetime region close to $r=0$ in which (i) the null-energy condition is violated, (ii) radial null geodesics are repelled from the core and (iii) curvature invariants are bounded from above. Physically, this may be interpreted as a spacetime region in which gravity acts as a repulsive force.
\item[c)] Geodesic completeness appears to be challenging to achieve together with regularity of curvature invariants. This is because the asymptotic behavior of the mass function near $r=0$ required for regularity results in a change of sign in $\frac{dr}{dv}$ for null geodesics. On the one hand, this implies that null geodesics no longer terminate at finite time $v$ and become future-complete. On the other hand, they can start at $r=0$ at finite time and there are thus past-incomplete null geodesics. This suggests that an extension of the spacetime to $r<0$ may be necessary. Due to the geodesics that start at $r=0$ at finite $v$, the spacetime does not admit a well-defined initial value problem.
\item[d)] Dynamical instabilities due to backreaction, are expected to occur, cf.~\cite{Carballo-Rubio:2024dca}. In the present case, this is because null geodesics are gravitationally lensed towards an attractor initially close to, then converging towards, the inner horizon. For spacetime with a horizon, the attractor is the boundary between the spacetime region in which the null energy condition holds and the region in which it is violated. Loosely speaking, these are spacetime regions with attractive and repulsive gravity and thus null geodesics are focused towards the boundary. Thus, one may expect a build-up of energy and thus an increasingly large backreaction at that boundary.\footnote{We caution that it depends on the spacetime dynamics whether a large backreaction results in an instability. In particular, any dynamics which regularizes the core of a black hole can be expected to contain a mechanism which limits the response of spacetime curvature to an increase in the energy-momentum-tensor, in other words, it limits the maximum effect that backreaction can have.}
\item[e)] For any choice of $r_{\rm NP}$, the spacetime is horizonless at very early times $v \gtrsim 0$ even for $\mu>\mu_{\rm c}$. Nevertheless, the spacetime region in which deviations from GR are large, is not accessible to observers at asymptotic infinity, because the attractor captures null geodesics. One may interpret this as a form of ``cosmic censorship of quantum gravity'', because the spacetime region in which quantum-gravity effects are sizable is dynamically shielded from far-away observers, and that includes the past-incomplete geodesics starting at $r = 0$ at finite $v$.
\item[f)] For $\mu <\mu_{\rm c}$, the classical spacetime contains a naked singularity. This singularity is regularized by the modification function. At the same time, the behavior of null geodesics that start at $(r=0, v=0)$ is unmodified in that they can reach asymptotic observers. Therefore, this case allows in principle an asymptotic observer to explore the spacetime region with large deviations from GR. It is a key question for future research, whether more realistic models of gravitational collapse share this feature.
\end{enumerate}

Our analysis motivates future work along the following lines.\\
First, continuing with the modified Vaidya spacetime, it is interesting to explore (timelike and null) geodesics in more detail, without the restriction to radial null geodesics; and exploring geodesic motion at $r<0$ can shed more light onto the question of geodesic completeness.\\
Second, if we can
extend the principled-parameterized approach to dynamical spacetimes with less symmetries, i.e., axisymmetric dynamical spacetimes such as the Kerr-Vaidya spacetime \cite{Senovilla_2015}, we can learn more about the universality of mass functions that render spacetimes with Killing vectors regular. However, this analysis might be complicated by the foliation-dependence of the apparent horizon in the Kerr-Vaidya spacetime \cite{Senovilla_2015,Dahal_2021}.\\
Furthermore, exploring whether the spacetime is unstable when backreaction is included and understanding the relevant timescales is important to determine whether the spacetimes we have explored may be viable despite the presence of an inner photon surface.

\section*{Acknowledgments}
\addcontentsline{toc}{section}{Acknowledgments}
We thank R.~Gold and A.~Platania for fruitful discussions and helpful suggestions. This work was supported by the Perimeter Institute for Theoretical Physics. Research at Perimeter Institute is supported in part by the Government of Canada through the Department of Innovation, Science and Economic Development and by the Province of Ontario through the Ministry of Colleges and Universities. This work is supported by a grant from VILLUM Fonden, no.~29405.

\newpage
\appendix
\appendixpage
\addappheadtotoc

\section{ZM invariants for the VKP model}
\label{app:ZMinvariants}

The 17 ZM polynomial curvature invariants --- built out of the Weyl tensor $C^{\mu \nu \rho \sigma}$, its dual $\overline{C}^{\mu\nu\rho\sigma}$ and the Ricci tensor --- for the VKP model with mass function Eq.~\eqref{eq:VKPmass} correspond to
\begin{align}
\label{eq:ZM-I1}
	I_{1} &=C_{\mu\nu\rho\sigma}C^{\mu\nu\rho\sigma}
	\\
	&=\frac{48 G_0^2 \mu^2 v^2}{r^6}, \nonumber\\ 
	I_{2} &=C_{\mu\nu\rho\sigma}\overline{C}^{\mu\nu\rho\sigma}
	\\
	&=0, \nonumber\\
	I_{3} &=C_{\mu\nu}^{\phantom{\mu\nu}\rho\sigma}C_{\rho\sigma}^{\phantom{\rho\sigma}\alpha\beta }C_{\alpha\beta }^{\phantom{\alpha\beta }\mu\nu}
	\\
	&=\frac{96 G_0^3 \mu^3 v^3}{r^9} = \frac{1}{2 \sqrt{3}}\, I_1^{3/2}, \nonumber\\
	I_{4} &=\overline{C}_{\mu\nu}^{\phantom{\mu\nu}\rho\sigma}C_{\rho\sigma}^{\phantom{\rho\sigma}\alpha\beta }C_{\alpha\beta }^{\phantom{\alpha\beta }\mu\nu}
	\\
	&=0, \nonumber\\
	I_{5} &=R
	\\
	&=0, \nonumber\\
	I_{6} &=R_{\mu}^{\phantom{\mu}\nu}R_{\nu}^{\phantom{\nu}\mu}
	\\
	&=0, \nonumber\\
	I_{7} &=R_{\mu}^{\phantom{\mu}\nu}R_{\nu}^{\phantom{\nu}\rho}R_{\rho}^{\phantom{\rho}\mu}
	\\
	&=0, \nonumber\\
	I_{8} &=R_{\mu}^{\phantom{\mu}\nu}R_{\nu}^{\phantom{\nu}\rho}R_{\rho}^{\phantom{\rho}\sigma}R_{\sigma}^{\phantom{\sigma}\mu}
	\\
	&=0, \nonumber\\
	I_{9} &=R^{\mu\nu}R^{\rho\sigma} C_{\mu\rho\nu\sigma}
	\\
	&=0, \nonumber\\
	\label{eq:I10}
	I_{10} &=R^{\mu\nu}R^{\rho\sigma} \overline{C}_{\mu\rho\nu\sigma}
	\\
	&=0, \nonumber\\
	I_{11} &=R^{\nu\rho}R_{\gamma\delta }\left(
		C_{\mu\nu\rho\sigma}C^{\mu \gamma\delta \sigma} 
		- \overline{C}_{\mu\nu\rho\sigma}\overline{C}^{\mu\gamma\delta \sigma}
	\right)
	\\
	&=0, \nonumber\\
	\label{eq:I12}
	I_{12} &= 2 R^{\nu\rho}R_{\gamma\delta }C_{\mu\nu\rho\sigma}\overline{C}^{\mu\gamma\delta\sigma}
	\\
	&=0, \nonumber\\
	I_{13} &=R_{\mu}^{\phantom{\mu}\gamma}R_{\gamma}^{\phantom{\gamma}\rho}R_{\nu}^{\phantom{\nu}\delta}R_{\delta}^{\phantom{\delta}\sigma}C^{\mu\nu}_{\phantom{\mu\nu}\rho\sigma}
	\\
	&=0, \nonumber\\
	I_{14} &=R_{\mu}^{\phantom{\mu}\gamma}R_{\gamma}^{\phantom{\gamma}\rho}R_{\nu}^{\phantom{\nu}\delta}R_{\delta}^{\phantom{\delta}\sigma}\overline{C}^{\mu\nu}_{\phantom{\mu\nu}\rho\sigma}
	\\
	&=0, \nonumber\\
	I_{15} &=\frac{1}{16}R^{\nu\rho}R_{\gamma\delta}\left(
		C_{\mu\nu\rho\sigma}C^{\mu\gamma\delta\sigma} 
		+ \overline{C}_{\mu\nu\rho\sigma}\overline{C}^{\mu\gamma\delta\sigma}
	\right)
	\\
	&= 0, \nonumber\\
	\label{eq:ZM-I16}
	I_{16} &=\frac{1}{32}R^{\rho\sigma}R^{\gamma\delta }C^{\mu\kappa\lambda\nu}\left(
		C_{\mu\rho\sigma\nu}C_{\kappa\gamma\delta\lambda} 
		+ \overline{C}_{\mu\rho\sigma\nu}\overline{C}_{\kappa\gamma\delta\lambda}
	\right)
	\\
	&=0, \nonumber\\
	I_{17} &=\frac{1}{32}R^{\rho\sigma}R^{\gamma\delta }\overline{C}^{\mu \kappa\lambda \nu}\left(
		C_{\mu\rho\sigma\nu}C_{\kappa\gamma\delta\lambda} 
		+ \overline{C}_{\mu\rho\sigma\nu}\overline{C}_{\kappa\gamma\delta\lambda}
	\right)\\
	&=0. \nonumber
	\label{eq:ZM-I17}
\end{align}
As a result, only the curvature invariants $I_1$ and $I_3$ (proportional to each other) do not vanish.

\section{Defining equation for the apparent horizon}
\label{app:DefEqApparentHorizon}
Here we derive the defining equation for the location of the apparent horizon and discuss its solutions.\\

Being a trapped surface, the location of the apparent horizon is determined by finding the null outgoing $\Theta_{\rm out}$ and null ingoing $\Theta_{\rm in}$ expansions such that
\be
\Theta_{\rm out} = 0,\quad \Theta_{\rm in} < 0.
\label{eq:Null_Expansions}
\ee
It has been shown in \cite{Dahal:2021vjr}\footnote{We recover the expansions of a generalized Vaidya spacetime by taking the limit $a \rightarrow 0$ of the null expansions in Kerr-Vaidya.} that the null expansions in a generalized Vaidya spacetime take the form
\be
\Theta_{\rm out} = \frac{1}{r} \left(1 - \frac{2 G_0 m[v,r]}{r}\right),\quad \Theta_{\rm in} = -\frac{2}{r}.
\ee
Hence, finding the location of the apparent horizon through the conditions in Eq.~\eqref{eq:Null_Expansions} amounts to satisfy
\be
1 - \frac{2 G_0 m[v,r]}{r} = 0,\quad r > 0.
\ee
We subsequently show how to arrive to the same conclusion following another method based on a change of coordinates at the level of the metric.\\

We start from the result derived in \cite{Nielsen:2005af,Faraoni:2013aba}, where the authors argue that any spherically symmetric line element can be written
in Painlev\'e-Gullstrand coordinates in the form\footnote{We use $v_1[r,t]$ instead of the notation $v[r,t]$ in \cite{Nielsen:2005af} to avoid confusion with the coordinate $v$.}
\bea
ds^2_{\rm PG} &=&g_{\rm tt, \, PG}\,dt^2 + 2 g_{\rm tr,\, PG}\,dt\, dr + dr^2 + r^2 d\Omega^2\nonumber\\
 &=&-\left(c[r,t]^2 - v_1[r,t]^2\right)\, dt^2 + 2 v_1[r,t]\, dt\, dr + dr^2 + r^2 d\Omega^2.\label{eq:PGcoordinates}
\eea
The apparent horizon is found from the expansions of the in- and outgoing radial null geodesics. In the above form of the line-element, it is simple to see where both expansions vanish. Radial null geodesics fulfill the condition
\be
0=-\left(c[r,t]^2 - v_1[r,t]^2\right)\, dt^2 + 2 v_1[r,t]\, dt\, dr + dr^2,
\ee
leading to
\be
\frac{dr}{dt} = - v_1[r,t] \pm c[r,t].
\ee
For $c[r,t]< v_1[r,t]$, outgoing null geodesics are no longer moving towards larger $r$, such that the apparent horizon is defined by the condition
\be
c[r,t]=v_1[r,t]. 
\ee
In turn, this condition can be rewritten in the form \cite{Faraoni:2013aba}
\be
g^{rr}_{\rm PG}=0.
\ee
We now show that if a coordinate transformation into ingoing Eddington-Finkelstein coordinates exists\footnote{\cite{Nielsen:2005af} argues  that any spherically symmetric metric can be written in the form Eq.~\eqref{eq:PGcoordinates}. Accordingly, starting from a spherically symmetric metric in Eddington-Finkelstein coordinates, the transformation must in principle exist, even if in practice it may take a form that is difficult to solve explicitly.}, then this condition implies $g^{rr}_{\rm EF}=0$.
Starting from Painlev\'e-Gullstrand coordinates, we make the transformation
\be
t=h[v,r],
\ee
with a function $h[v,r]$ that is to be determined. Omitting the explicit dependence of the transformation on coordinates, we arrive at
\bea
ds^2&=&g_{\rm tt, \, PG} \left(h^{(1,0)}\right)^2 dv^2 + \left(2 g_{\rm tt,\,PG}h^{(1,0)}h^{(0,1)}+2g_{\rm tr,\,PG}h^{(1,0)} \right) dr\, dv\nonumber\\
&{}&+ \left(g_{\rm tt,\, PG}h^{(0,1)}+2g_{\rm tr,\,PG}h^{(0,1)}+1 \right)dr^2+r^2\, d\Omega^2,
\eea
where $h^{(1,0)} \equiv \frac{\partial h[v,r]}{\partial v}$ and similarly $h^{(0,1)} \equiv \frac{\partial h[v,r]}{\partial r}$.
To correspond to the form Eq.~\eqref{eq:Vaidyametric}, we require
\bea
2&=&2 g_{\rm tt,\,PG}h^{(1,0)}h^{(0,1)}+2g_{\rm tr,\,PG}h^{(1,0)},\\
0&=&g_{\rm tt,\, PG}h^{(0,1)}+2g_{\rm tr,\,PG}h^{(0,1)}+1,
\eea
which, in turns, leads to
\bea
h^{(1,0)} &=& \frac{\partial h[v,r]}{\partial v} = \frac{g_{\rm tt,\,PG} + 2 g_{\rm tr,\,PG}}{g_{\rm tt,\,PG} \left(g_{\rm tr,\,PG} - 1\right) + 2 g_{\rm tr,\,PG}^2}, \nonumber\\
h^{(0,1)} &=& \frac{\partial h[v,r]}{\partial r} = \frac{-1}{g_{\rm tt,\,PG} + 2g_{\rm tr,\,PG}}.
\eea
Assuming that the two conditions can be solved to provide an $h[v,r]$, we can relate $g^{rr}_{\rm PG}$ to $g^{rr}_{\rm EF}$ by
\be
g^{\rm rr}_{\rm EF} = -g_{\rm tt, \, EF} = - g_{\rm tt,\, PG}\left(h^{(1,0)} \right)^2= g^{\rm rr}_{\rm PG} \left(g_{\rm tt,\, PG}- g_{\rm tr,\, PG} \right) \left(h^{(1,0)} \right)^2.
\ee
Accordingly, $g^{\rm rr}_{\rm PG}=0$ implies $g^{\rm rr}_{\rm EF}=0$.

We now evaluate the condition $g^{\rm rr}_{\rm EF}=0$ for the generalized Vaidya metric in Eq.~\eqref{eq:GeneralisedVaidyametric} with upgraded mass \eqref{eq:upgradedmassfunction}: 
\bea
g^{rr}_{\rm EF}&=&g_{\rm vv,\, EF},\nonumber\\
&=& 1- \frac{2G_0}{r}m[v,r],\nonumber\\
&=& 1- \frac{2G_0}{r}m[v]f_{\rm NP}\left[I_1 r_{\rm NP}^4 \right].
\label{eq:EquationApparentInnerHorizons}
\eea
As we have discussed in Sec.~\ref{subsubsec:simplicity}, $f_{\rm NP} \leq 1$ holds for all $r$. This implies that the apparent horizon lies at smaller radii, i.e., the black hole is more compact.\\

Further, the condition $g^{rr}_{\rm EF}=0$ generically has two real solutions. The first is the shifted apparent horizon, the second is a new, inner apparent horizon, which occurs, because $f_{\rm NP} \rightarrow 0$ for small $r$. If $r_{\rm NP}$ is large enough, the two horizons merge; for larger $r_{\rm NP}$, the spacetime is horizonless, because the solutions to $g^{rr}=0$ are all complex in this regime.\\
These results mirror similar results in the stationary limit with a constant ADM mass $m$. This is because in the stationary case, $g^{rr}=0$ is the condition for the event horizon. Upon the identification $m= \mu \bar{v}$, the two equations become identical and thus the time-dependent case must agree with the stationary case for each fixed value $\bar{v}$.

\section{Generalised ZM invariants}
\label{app:GeneralisedZMinvariants}
The 17 ZM polynomial curvature invariants for a generalized mass function $m[v,r]$ with dependencies on both advanced time and radial coordinate correspond to
\begin{align}
\label{eq:GenZM-I1}
	I_{1} &=C_{\mu\nu\rho\sigma}C^{\mu\nu\rho\sigma}
	\\
	&=\frac{4 G_0^2}{3r^6} \left(6 m[v,r] + r \left(-4 m^{(0,1)}[v,r]+ r m^{(0,2)}[v,r] \right) \right)^2,\\
	I_{2} &=C_{\mu\nu\rho\sigma}\overline{C}^{\mu\nu\rho\sigma}
	\\
	&=0,\\
	I_{3} &=C_{\mu\nu}^{\phantom{\mu\nu}\rho\sigma}C_{\rho\sigma}^{\phantom{\rho\sigma}\alpha\beta }C_{\alpha\beta }^{\phantom{\alpha\beta }\mu\nu}
	\\
	&=\frac{4 G_0^3}{9 r^9} \left(6 m[v,r] + r \left(-4 m^{(0,1)}[v,r]+ r m^{(0,2)}[v,r] \right) \right)^3 = \frac{1}{2 \sqrt{3}}\, I_1^{3/2},\\
	I_{4} &=\overline{C}_{\mu\nu}^{\phantom{\mu\nu}\rho\sigma}C_{\rho\sigma}^{\phantom{\rho\sigma}\alpha\beta }C_{\alpha\beta }^{\phantom{\alpha\beta }\mu\nu}
	\\
	&=0,\\
	I_{5} &=R
	\\
	&=G_0 \cdot \frac{4 m^{(0,1)}[v,r]+ 2 r \,m^{(0,2)}[v,r]}{r^2},\\
	I_{6} &=R_{\mu}^{\phantom{\mu}\nu}R_{\nu}^{\phantom{\nu}\mu}
	\\
	&=G_0^2 \cdot\frac{8 \left(m^{(0,1)}[v,r]\right)^2+ 2 r^2\left( m^{(0,2)}[v,r]\right)^2}{r^4},\\
	I_{7} &=R_{\mu}^{\phantom{\mu}\nu}R_{\nu}^{\phantom{\nu}\rho}R_{\rho}^{\phantom{\rho}\mu}
	\\
	&=G_0^3 \cdot \frac{16 \left(m^{(0,1)}[v,r]\right)^3+ 2 r^3\left( m^{(0,2)}[v,r]\right)^3}{r^6},\\
	I_{8} &=R_{\mu}^{\phantom{\mu}\nu}R_{\nu}^{\phantom{\nu}\rho}R_{\rho}^{\phantom{\rho}\sigma}R_{\sigma}^{\phantom{\sigma}\mu}
	\\
	&=G_0^4 \cdot \frac{32 \left(m^{(0,1)}[v,r]\right)^4+ 2 r^2\left( m^{(0,2)}[v,r]\right)^4}{r^8},\\
	I_{9} &=R^{\mu\nu}R^{\rho\sigma} C_{\mu\rho\nu\sigma}\\
	&= \frac{2 G_0^3}{3 r^7} \left(-2 m^{(0,1)}[v,r]+r \,m^{(0,2)}[v,r]\right)^2 \nonumber\\
	&\quad\cdot \left(6 m[v,r] + r \left(-4 m^{(0,1)}[v,r]+ r m^{(0,2)}[v,r]\right)\right),\\
	\label{eq:GenZM-I10}
	I_{10} &=R^{\mu\nu}R^{\rho\sigma}\overline{C}_{\mu\rho\nu\sigma}\\
	&=0,\\
	I_{11} &=R^{\nu\rho}R_{\gamma\delta }\left(
		C_{\mu\nu\rho\sigma}C^{\mu \gamma\delta \sigma} 
		- \overline{C}_{\mu\nu\rho\sigma}\overline{C}^{\mu\gamma\delta \sigma}
	\right)\\
	&=\frac{I_9 \sqrt{I_1}}{\sqrt{3}},\\
	\label{eq:GenZM-I12}
	I_{12} &= 2 R^{\nu\rho}R_{\gamma\delta }C_{\mu\nu\rho\sigma}\overline{C}^{\mu\gamma\delta\sigma}
	\\
	&=0,\\
	I_{13} &=R_{\mu}^{\phantom{\mu}\gamma}R_{\gamma}^{\phantom{\gamma}\rho}R_{\nu}^{\phantom{\nu}\delta}R_{\delta}^{\phantom{\delta}\sigma}C^{\mu\nu}_{\phantom{\mu\nu}\rho\sigma}
	\\
	&=\frac{G_0^5}{\sqrt{3} r^8} \left(-4 \left(m^{(0,1)}[v,r]\right)^2+ r^2 \left(m^{(0,2)}[v,r]\right)^2 \right)^2 \sqrt{I_1},\\
	I_{14} &=R_{\mu}^{\phantom{\mu}\gamma}R_{\gamma}^{\phantom{\gamma}\rho}R_{\nu}^{\phantom{\nu}\delta}R_{\delta}^{\phantom{\delta}\sigma}\overline{C}^{\mu\nu}_{\phantom{\mu\nu}\rho\sigma}
	\\
	&=0,\\
	I_{15} &=\frac{1}{16}R^{\nu\rho}R_{\gamma\delta}\left(
		C_{\mu\nu\rho\sigma}C^{\mu\gamma\delta\sigma} 
		+ \overline{C}_{\mu\nu\rho\sigma}\overline{C}^{\mu\gamma\delta\sigma}
	\right)
	\\
	&= \frac{1}{16}I_{11},\\
	\label{eq:GenZM-I16}
	I_{16} &=\frac{1}{32}R^{\rho\sigma}R^{\gamma\delta }C^{\mu\kappa\lambda\nu}\left(
		C_{\mu\rho\sigma\nu}C_{\kappa\gamma\delta\lambda} 
		+ \overline{C}_{\mu\rho\sigma\nu}\overline{C}_{\kappa\gamma\delta\lambda}
	\right)
	\\
	&=\frac{1}{8 \sqrt{3}} I_{11} \sqrt{I_1},\\
	I_{17} &=\frac{1}{32}R^{\rho\sigma}R^{\gamma\delta }\overline{C}^{\mu \kappa\lambda \nu}\left(
		C_{\mu\rho\sigma\nu}C_{\kappa\gamma\delta\lambda} 
		+ \overline{C}_{\mu\rho\sigma\nu}\overline{C}_{\kappa\gamma\delta\lambda}
	\right)\\
	&=0,
	\label{eq:GenZM-I17}
\end{align}
where $m^{(0, n)}[v,r] = \frac{\partial^n m[v,r]}{\partial r^n}$ and no partial derivatives with respect to the advanced time $v$ appear.

\bibliography{References}
	
\end{document}